\documentclass[final,5p,times,twocolumn]{elsarticle}

\usepackage{url}
\usepackage{bbm}
\usepackage{amsmath,amsfonts} 
\usepackage{float}
\usepackage{graphicx}
\usepackage{algorithm}
\usepackage{algpseudocode}
\usepackage{siunitx}
\usepackage[labelfont=bf]{caption}
\usepackage{subcaption}
\usepackage{tikz}
\usepackage{pgfplots}
\usepackage{epstopdf}
\usepackage{multirow}
\usepackage[font=scriptsize]{caption}
\usepackage{soul}
\usepackage{color, colortbl}
\usepackage{array}
\usepackage[T1]{fontenc}
\usepackage[utf8]{inputenc}
\usepackage{mathptmx}
\usepackage{listings}
\usepackage{adjustbox}
\usepackage{makecell}
\usepackage{tcolorbox}
\usepackage{dsfont}
\usepackage{tabularx}

\usepackage{enumitem}
\usepackage{hyperref}
\usepackage{pifont}

\DeclareMathOperator*{\argmin}{argmin}

\def\BibTeX{{\rm B\kern-.05em{\sc i\kern-.025em b}\kern-.08em
    T\kern-.1667em\lower.7ex\hbox{E}\kern-.125emX}}

\begin{document}

\begin{frontmatter}

\title{Explaining Intrusion Alert Decisions of Deep Learning-based Network Intrusion Detection Systems for Security Analysts}

\author{Ayush Kumar, Vrizlynn L.L. Thing} 

\affiliation{organization={Cyber Security Strategic Technology Centre, ST Engineering},
            addressline={600 W Camp Rd}, 
            city={Singapore},
            postcode={797654}, 
            state={Singapore},
            country={Singapore}}

\begin{abstract}
While Deep Learning-based Network Intrusion Detection Systems (DL-based NIDS) have recently been significantly explored and shown superior performance, they are insufficient to actively respond to the detected intrusions (security alerts) due to the semantic gap between the alerts and actionable interpretations. Furthermore, their high error costs make security analysts unwilling to respond solely based on the alerts generated by them. The root cause of these drawbacks can be traced to the lack of explainability of alerts generated by DL-based NIDS. Although some methods have been developed to explain DL-based NIDS (e.g., xNIDS \cite{xnids}), they cannot explain alerts generated by such NIDS in enterprise settings in a way that is understandable by security analysts. Further, those explanation methods are incapable of handling feature dependencies in network traffic which are more complex than what has been assumed during their implementation (features can only be divided into exclusive, equally important groups). 

In this paper, we present \textsf{EXP-SEC}, a novel framework which can explain the intrusion detection decisions of DL-based NIDS (which lead to security alerts) in a way that is aligned with the domain knowledge of analysts working in Security Operation Centres (SOC). We highlight the following features of our framework: (1) a forensic module that isolates the suspect packets/flow which likely caused an alert (2) an explanation module which can handle much more complex feature dependencies in network traffic than existing methods (features can be divided into overlapping groups and some groups are more important than others), and (3) a multi-stage mapping module which translates the feature/group-based explanations generated by explanation module to domain-specific explanations suitable for processing by security analysts. We evaluate EXP-SEC with state-of-the-art DL-based NIDS and our evaluation results show that EXP-SEC outperforms xNIDS (existing best performing explanation framework) in terms of group-level and overlap-aware explanation utility metrics while performing similarly in terms of conventional feature-level metrics such as descriptive accuracy, sparsity and stability. Moreover, taking the case of a state-of-the-art DL-based NIDS, we demonstrate the security analyst-friendly explanation format generated by EXP-SEC.

\end{abstract}



\end{frontmatter}

\section{Introduction}
\label{intro}

Securing networks against intrusion is paramount, and network intrusion detection systems (NIDS) play a pivotal role in achieving this. Traditionally, NIDSs have relied heavily on signature-based methods \cite{bro, snort}- essentially matching attack patterns to known signatures. However, these approaches demand significant manual effort from experts to create and maintain those signatures. Moreover, they fall short when it comes to identifying novel, previously unseen attacks. With the introduction of deep learning (DL)-based NIDSs which leverage DL models to enhance existing defenses, impressive performance and accuracy has been achieved \cite{dd-nids, MLIDissues} and they have been found to perform better than conventional machine learning (ML) algorithms \cite{kitsune}.
In fact, unsupervised DL-based NIDSs can even raise alerts for hitherto unknown attacks. Thanks to these advantages and the rapid evolution of DL techniques, DL-based NIDSs have made significant strides in recent years.

Despite the promise of DL-based NIDSs, there is a noticeable hesitance among network security experts and practitioners when it comes to deploying them in real-world production environments \cite{dos-n-donts-ml}. The crux of the issue lies in the black-box nature of many of these proposed solutions. Essentially, their inner workings (how and why they make decisions) remain elusive, unlike simpler but less effective rule-based approaches that security professionals are accustomed to. Security analysts rely on DL-based NIDSs for critical security alerts. However, without clear explanations for the alerts generated by these underlying DL models, it becomes difficult for the analysts to trust them. 
Moreover, mis-detections generated by DL-NIDS require analysts to spend a significant amount of time and effort on troubleshooting, resulting in high error costs. Due to the low explainability of state-of-the-art DL-based NIDSs \cite{ayush-iot-nids-explain}, it is difficult to understand why they generate certain alerts, which makes troubleshooting detection errors difficult and lowers the confidence of corresponding intrusion responses.

Several methods \cite{warnecke, maonan, cade, deepaid} have been developed to explain results from deep learning models. Although it is difficult to explain the entire prediction process globally, existing methods provide a local explanation for a specific input sample and the decision on it by highlighting the most important features [55]. For example, Gradients \cite{grad}, IG \cite{ig}, LRP \cite{lrp}, LIME \cite{lime}, and SHAP \cite{shap} are used to explain image classifiers and Natural Language Processing (NLP) models. LEMNA \cite{lemna}, which is specifically designed for deep learning-based security applications, can explain binary code analysis tools and malware detection tools. xNIDS \cite{xnids} can explain DL-based NIDS and is designed to address the limitations of explanation methods published earlier (IG, LRP, LIME, SHAP, LEMNA) by approximating and sampling the history inputs and capturing the feature dependencies with sparse group lasso. xNIDS has been showed to outperform those explanation methods across multiple metrics.

However, xNIDS also suffers from limitations. It used standard SGL (Sparse Group Lasso) penalty-based optimization to find feature coefficients (contribution to DL-based NIDS decision) which applies the same static penalty factor ($\lambda$) to all feature groups as well as all individual features. Network traffic features can usually be grouped, e.g., by protocol type (TCP, UDP) but one group may have more predictive power than another. SGL provides no way to adjust the penalty for different groups based on their predictive power. Further, SGL does not support overlapping of features among groups which can also occur in network traffic features. If we force SGL to run on overlapping groups, it leads to a logical contradiction within the optimization loop wherein an active and an inactive group share a feature. Either the coefficient for shared feature has to be set to zero, affecting predictive power of active group, or the inactive group should be included in the final model, affecting group-level sparsity.  

In this work, we present \textsf{EXP-SEC}, a new framework that explains the intrusion detection decisions of DL-based NIDSs (which lead to security alerts) in alignment with domain-specific knowledge so that the explanations are understandable and actionable for security analysts. Our proposed explanation method addresses the aforementioned limitations of xNIDS by: (1) finding packets/flows in the past input traffic stream that most likely led to a security alert, and (2) capturing feature dependencies of the structured traffic data with feature groups (which can have overlapping features and some of which can be assigned more importance than others based on predictive power) and adaptive sparse group lasso \cite{yaglm}. The feature-based explanation thus generated is mapped to security domain-specific explanation using a multi-stage pipeline consisting of feature extraction framework de-construction, a correlation engine and a semantic reasoning engine. To evaluate EXP-SEC, we apply it to state-of-the-art DL-NIDSs such as: (1) the autoencoder-based Kitsune \cite{kitsune} and (2) the Recurrent Neural Networks (RNN) based RNN-IDS \cite{rnn-ids}. Our evaluation results show EXP-SEC performs similarly as the existing best performing explanation framework (xNIDS) in terms of conventional explanation utility metrics such as descriptive accuracy, sparsity, stability and runtime. Further, EXP-SEC outperforms xNIDS in terms of new group-level and overlap-aware metrics. It should be noted that the underlying penalization technique used for explanation coefficient generation in EXP-SEC (adaptive sparse group lasso with overlapping groups) subsumes the technique used in xNIDS (sparse group lasso) and is therefore, more generalized.

The main contributions of our work are as follows:
\begin{itemize}
	\item We design a novel explanation framework, \textsf{EXP-SEC}, dedicated to explaining intrusion detection decisions of DL-based NIDS to security analysts. Our framework isolates the packets/flows most likely behind the alert (can be used to re-train the NIDS) and captures the complex network traffic feature dependencies with adaptive sparse group lasso (with overlapping groups) penalization.
	\item We present a multi-stage mapping methodology as part of EXP-SEC to translate the feature-based explanations obtained earlier to domain-specific explanations for security analysts.
	\item We evaluate EXP-SEC with state-of-the-art DL-based NIDSs, demonstrate the effectiveness of its underlying explanation technique, and use a case study to show how it can help security analysts understand alerts generated by a DL-based NIDS during different attack scenarios.
\end{itemize}
 

\section{Background}
\label{background}
In this section, we briefly explore the various types of explanation methods present in academic literature and why they fall short when it comes to explaining real-world NIDS alerts to security analysts. 



\subsection{Overview of Existing Explanation Methods}
All explanation methods for ML/DL models can be broadly categorized into \textit{black-box} and \textit{white-box} explanations.

\textbf{Black-box Explanations} These methods function within a black-box context, assuming zero knowledge about the inner workings of the underlying ML model and its associated parameters. These methods prove invaluable when direct access to the ML model is not available, for instance, during remote audits of a learning-based application. These black-box methods approximate the underlying ML model’s prediction function $f_N$, allowing them to estimate how various dimensions of the input vector $x$ affect a given prediction. While they hold promise for explaining deep learning models, their effectiveness can be impeded by the black-box environment, sometimes leaving out crucial insights embedded in the ML model’s parameters and architecture.

\textbf{White-box Explanations} These methods work under the premise that we have full knowledge of all the parameters of the ML model being tested. Armed with this information, these methods bypass approximations and can directly compute explanations for the prediction function, $f_N$. The same ML model is often used to generate both predictions and explanations, assuming that the model is available on hand. This assumption holds true for stand-alone automated systems such as those that analyse binaries, detect malware or discover software vulnerabilties. However, it is worth noting that some white-box methods are customized for specific neural network architectures deployed in the fields of computer vision/speech processing/natural language processing and may not perform well for other architectures.

DL-based NIDS products are expected to be used by end users/customers who do not have access to the ML model and its parameters underlying the NIDS. This is to maintain confidentiality of the NIDS implementation by the cybersecurity services provider. Therefore, we assume that the DL-based NIDSs being explained in this work are available to us as a black-box system only and we use XAI techniques designed to explain such black-box ML systems. 
In what follows, we show a few representative explanation methods and analyze their limitations.

\textbf{Overview of LIME}: Ribeiro et al. \cite{lime} introduced LIME, one of the first black-box methods for explaining neural networks that is further extended by SHAP \cite{shap}. Both works are motivated by the need to allow users to be able to trust ML models as well as their predictions. LIME is designed to explain the predictions of any classifier or regressor in a faithful way, by approximating it locally with an interpretable model. 
LIME aims at approximating the decision function, $f$ of the model being explained by solving the following optimization problem:
\begin{equation}
\underset{g\in \mathcal{G}}{arg min} \ \mathcal{L}(f, g, \pi_x),
\end{equation}
where $\mathcal{L}(f, g, \pi_x)$ is the fidelity function which is a measure of how unfaithful the interpretable model $g$ is in approximating $f$ in the locality defined by $\pi_x$, $\mathcal{G}$ is the set of all linear functions and $\pi_x(z)$ is the proximity measure between an instance $z$ to $x$, so as to define the locality around $x$ and $x$ is the input to $f$.

In order to learn the local behavior of $f$ as the interpretable inputs vary, LIME approximates $\mathcal{L}(f, g, \pi_x)$ by drawing samples, weighted by $\pi_x$. The instances around $x$ are sampled by drawing nonzero elements of $x$ uniformly at random. Next, LIME creates a series of $l$ perturbations of $x$, denoted as $\tilde{x_1}$, $\dots$, $\tilde{x_l}$ by setting entries in the vector $x$ to $0$ randomly. The method then proceeds by predicting a label $f(\tilde{x_i}) = \tilde{y_i}$ for each $\tilde{x_i}$ of the $l$ perturbations. This strategy enables the method to approximate the local neighborhood of $f$ at the point $f(x)$. LIME approximates the decision boundary by a weighted linear regression model,
\begin{equation}
\mathcal{L}(f, g, \pi_x) = \sum_{i=1}^l \pi_x(\tilde{x_i})(f(\tilde{x_i})-g(\tilde{x_i}))^2 .
\end{equation}

\textbf{Overview of SHAP}: SHAP \cite{shap} (SHapley Additive exPlanations) subsumes additive feature attribution methods such as LIME and uses the following forms for $\mathcal{L}(f, g, \pi_x)$ and $\pi_x$ under the SHAP kernel method, which is shown to recover Shapley Values \cite{shapley-val} when solving the regression:
\begin{eqnarray*}
\mathcal{L}(f, g, \pi_{x'}) = \sum_{z' \in Z} [h_x^{-1}(z')-g(z')]^2 \pi_{x'}(z'), \\
\pi_{x'}(z') = \frac{M-1}{(M \text{choose} |z'|)|z'|(M-|z'|)} 
\end{eqnarray*}
where $x'$ is the simplified input which maps to the original input $x$ through a mapping function $x = h_x(x')$, $M$ is the number of simplified input features, $z' \in {\{0,1\}}^M$, and $|z'|$ is the number of non-zero elements in $z'$. 

Shapley values are a concept from game theory where the features act as players under the objective of finding a fair contribution of the features to the payout, in this case the prediction of the model. Different from TRUSTEE, Shapley values are used for local interpretation. In other words, it is used to interpret how a prediction result is reached for a specific data sample or subset of samples. Specifically, Shapley values can tell how each feature contributes to the predicted results. A positive Shapley value means that the feature’s value pushes the classification result toward being malicious, and a negative Shapley value does the contrary. 

\textbf{Overview of LEMNA}: LEMNA \cite{lemna} is a black-box method specifically designed for security applications. Given an input data instance $x$ and a classifier such as an RNN, this method aims to identify a small set of features that have key contributions to the classification of $x$. This is done by generating a local approximation of the target classifier’s decision boundary near $x$. It uses a mixture regression model (to approximate locally non-linear decision boundaries) enhanced by fused lasso (to handle correlated features) for approximation. The mixture regression model is a weighted sum of $K$ linear regression models:
\begin{equation}
f(x) = \sum_{j=1}^K \pi_j(\beta_j \cdot x + \epsilon_j).
\end{equation}
The parameter $K$ specifies the number of models, the random variables $\epsilon = (\epsilon_1, \dots , \epsilon_K)$ originate from a normal distribution $\epsilon_i \sim \mathcal{N}(0, \sigma)$ and $\pi = (\pi_1, \dots, \pi_K)$ holds the weights for each model. The variables $\beta_1, \dots , \beta_K$ are the regression coefficients and can be interpreted as $K$ linear approximations of the decision boundary near $f(x)$.

\textbf{Overview of xNIDS}: xNIDS is a framework that explains DL-based NIDS predictions and uses its explanation results to generate actionable defense rules. First, it heuristically finds a small number of history inputs that lead to a detection score in the vicinity of the original score and then samples in the vicinity of each of the history inputs by: (1) assigning larger weights to the latest history inputs and (2) shifting the synthesized samples towards the latest history inputs by weighted random sampling. Further, it captures the feature dependencies of the structured data for an explanation by first dividing the features into several groups based on their correlations and then applying a sparse group lasso to the feature groups to achieve a sparse explanation at the feature group and individual feature levels. The regression problem used in sparse group lasso is modelled as follows:

\begin{multline}
\label{eq-slass-opt}
	\argmin_{\beta} \left\{ {||f-g||}_2^2 + (1-\alpha)\lambda\sum_{q=1}^Q \sqrt{p_q}{||\beta_q||}_2 \right. \\
	\left. + \alpha\lambda{||\beta||}_1 \right\}
\end{multline}
where $\beta_q$ is a vector, which contains the coefficients for the features in $q^{th}$ group; $p_q$ is the size of $q^{th}$ group; $\beta = (\beta_1 \dots \beta_Q)$; and $\alpha \in [0,1]$ a convex combination of the lasso and group lasso penalties. To achieve group level sparsity, the expression $\sum{||\beta_q||}_2$ is minimized, namely excluding more groups by making ${||\beta_q||}_2 = 0$. To achieve feature-level sparsity, the expression ${||\beta||}_1$ is minimized, excluding more features by making ${||\beta_i||}_1 = 0$.

\subsection{Limitations of Existing Explanation Methods} 
LIME explanations are unstable since it draws random perturbations around the instance to learn the local surrogate and this sampling is stochastic, not suitable for domains with high feature correlation because it perturbs features independently, and assume that the decision boundary within the local neighbourhood of a data point is roughly linear which might not hold true. SHAP suffers from exponential computational complexity since calculating exact Shapley values requires evaluating the model prediction across all possible feature subsets which grows exponentially with the dimensions of the dataset, and it assumes independent features to approximate conditional expectations efficiently which fails in the case of highly correlated features. LEMNA relies heavily on the proper tuning of its mixture component count and fused lasso thresholds which makes it sensitive to poorly tuned hyper-parameters and its performance drops when applied to non-sequential tabular structures or graph-based relationships, as its internal regularization relies on ordered or structured dependencies between adjacent features. 

xNIDS fails to incorporate the flexibility to adjust the importance given to group-level and individual feature-level sparsity which is important when it comes to network traffic features (e.g., grouping network traffic features by protocol type where one protocol feature group may have more predictive power than another). (1) It uses SGL which applies the exact same group penalty $\lambda$ to a massive, completely irrelevant group of noise features as it does to a highly critical group. If we increase the penalty to block out the irrelevant groups, we can inadvertently damage the coefficient estimation within valid groups. (2) When features are grouped, but a few specific features or groups have massive true effect sizes while others have very small but non-zero effects, standard SGL struggles. To eliminate the vast number of true zero noise features, SGL forces a high penalty. This heavily suppresses the signal of most critical, dominant predictors, leading to underestimation of their true impact and poor predictive calibration. (3) Consider a scenario where features within a group are highly correlated (collinear), but only one or two features in that group are actually driving the outcome, while the rest are irrelevant. Due to the strong correlation, standard SGL struggles to separate the true signal from the correlated noise within the active group. It will either erroneously include the entire group without internal sparsity (behaving like a pure Group Lasso) or randomly select among the collinear features due to the unstable nature of the $L_1$ penalty under collinearity.

Further, xNIDS fails to account for the existence of overlapping features between feature groups which is a possible occurrence in network traffic features as explained in Section \ref{feat-grps}. Standard SGL is mathematically built on the assumption that groups form a strict partition, i.e., every feature belongs to exactly one group, and the intersection between any two groups is empty. If we force standard SGL to run on overlapping groups, it introduces three major problems: (1) The standard group lasso penalty encourages entire blocks of coefficients to be set to exactly zero. If a feature $x_i$ lies in the intersection of an active group ($G_1$) and an inactive group ($G_2$), a severe logical contradiction occurs within the optimization loop. To make $G_2$ sparse (inactive), the algorithm must set all coefficients in $G_2$ to zero, which includes $x_i$ and harms the predictive power of the active group $G_1$, or it unnecessarily forces $G_2$ to enter the model (affecting group-level sparsity) just to keep $x_i$ alive. (2) If a feature appears in 3 different groups, its coefficient $\beta_i$ is factored into three separate $L_2$ norm calculations, plus the global $L_1$ penalty which causes compounded penalization. (3) Standard SGL relies on proximal gradient descent methods (such as FISTA) or coordinate descent. For partitioned groups, the proximal operator of the group penalty has a closed-form solution because the sub-gradients of different groups are completely independent. When groups overlap, the sub-gradients become coupled, the proximal operators for individual groups no longer commute and it is no longer possible to solve the proximal step for each group independently in a single pass. The solver algorithm will either fail to converge, yield mathematically incorrect projections, or require computationally gruelling iterative inner loops.	  

\section{Proposed Explanation Method}
\begin{figure*}[t]
\centering
\includegraphics[scale=0.3]{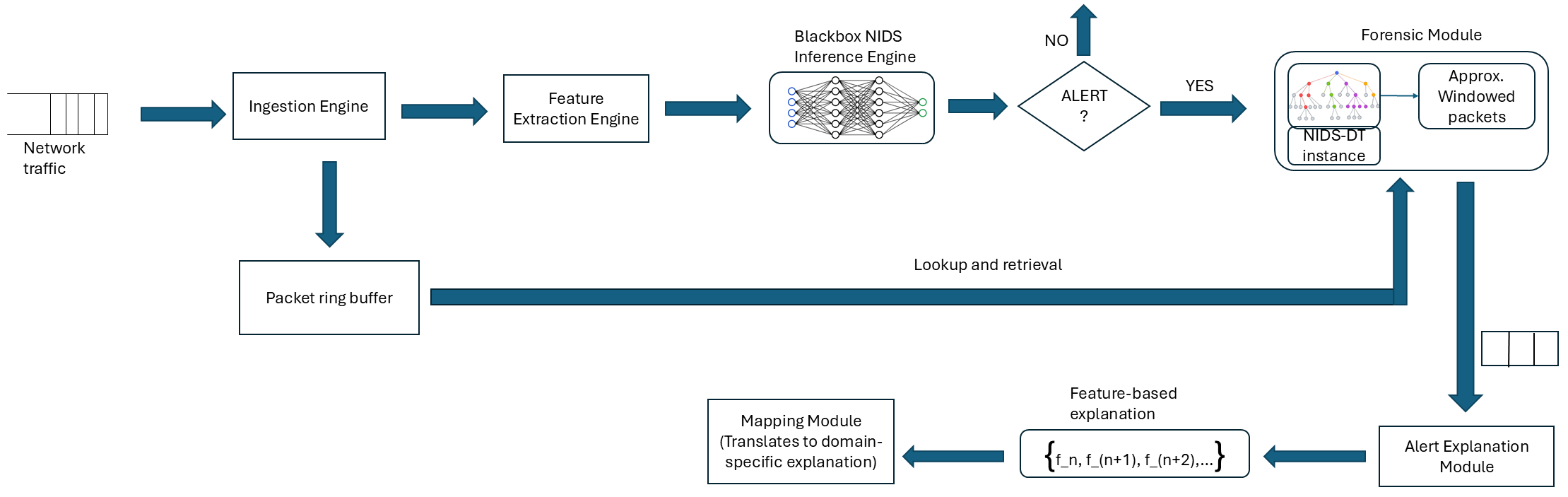}
\caption{EXP-SEC System Architecture. The continuous packet traffic is input to an \textit{Ingestion Engine} which duplicates the packet stream, one of which goes to the \textit{Feature Extraction Engine} and the other one to the \textit{Packet Ring Buffer}. The \textit{NIDS Inference Engine} processes the incoming feature vectors from feature extraction engine. When an alert is received, the \textit{Forensic Module} looks up and pulls the matching suspect packet traffic from the packet ring buffer. It then approximates the relevant windowed packets and send them to the security analysts while the features extracted from windowed packets are sent to \textit{Alert Explanation Module}. The explanation module generates the feature-level explanation which is translated to domain-specific explanation by the \textit{Mapping Module}.}
\label{exp-sec-arch}
\end{figure*}

State-of-the-art explanation methods are not designed as per the requirements of security analysts working with DL-based NIDSs deployed in real-world enterprise IT environments. Those explanation methods either provide local explanations for a blackbox model acting on a subset of data samples (e.g., SHAP, LIME) or global explanations for a blackbox model trained on an extensive dataset. However, security analysts deal with NIDSs operating with continuous traffic and need explanations for intrusion alerts raised by those NIDSs in terms of domain-specific explanation (e.g., attack tactic, threat indicator) for the alert decision and the traffic samples which caused the alert. Keeping those requirements in mind, we have designed EXP-SEC, our proposed explanation framework whose architecture is shown in Fig. \ref{exp-sec-arch}.

\subsection{Notation and Design Goals}
\subsubsection{Notation}
We use bold lowercase letters, such as $x$, to represent a vector and bold capital letters, such as $X$, to represent a matrix or a sequence of vectors. We denote by $X_{t,k}$ or $(x_t, \dots, x_{t-k})$ the windowed packet sequence of length $k$. An DL-NIDS is denoted by $f(X_{t,k}) \in \mathds{R}$. Denote by $g(X_{t',m}) \in \mathds{R}$ a local approximation to $f(X_{t,k})$. We define the explanation result as $e$, which contains the desired important features. We denote by $||\cdot||$ the $l_1$ and ${||\cdot||}_2$ the $l_2$ norm of a vector. Denote by $L(\cdot)$ the loss function that measures how faithful one model locally approximates another. $\phi(\cdot)$ denotes the sparsity of a vector.
\subsubsection{Our Goal}
Given an input $x_t$, our explanation method needs to find relevant inputs $X_{t',m}$ and a high-fidelity, sparse, complete, and stable explanation result $e$. In our design, $e$ is determined by the parameters $\beta$ of $g(\cdot)$. To this end, we formulate our explanation method by the following:

\begin{equation}
\label{eq-slass-opt}
	\argmin_{g} \biggl\{\underbrace{L(f,g)}_{\text{Fidelity}} + \underbrace{\lambda\cdot\phi(\beta)}_{\text{Sparsity}}\biggr\} \textit{s.t.} \biggl\{\underbrace{||f(x_t,X_{t',m})-y_t||<\delta}_{\text{Relevant windowed input}}\biggr\}
\end{equation}

\subsection{Suspicious Packet Traffic Isolation}
\label{sus-pkt-isol}
The continuous packet traffic is input to an \textit{Ingestion Engine} which captures packets at the line-rate without dropped frames. Every incoming packet is instantly tagged with a metadata header containing a Global Unique Identifier ($Pkt\_ID$), typically a combination of a monotonic nanosecond timestamp and a hash of the packet tuple. The packet stream is then immediately duplicated, such that the $Header + Payload$ metadata goes to the \textit{Feature Extraction Engine} and the raw packets go straight to the \textit{Packet Ring Buffer}. The feature extraction engine builds the windowed features required by the NIDS from the incoming network packets/flows as per the feature extraction framework (e.g., \textit{AfterImage} for Kitsune). 

The \textit{Packet Ring Buffer} is a lock-free circular ring buffer which stores the raw packets indexed strictly by their $Pkt\_ID$. It operates on a sliding TTL (Time-To-Live) basis wherein for example, if the NIDS window maximum look-back is $t$ minutes, the ring buffer overrides data older than $2 \times t$ minutes to maintain a constant memory footprint. The \textit{NIDS Inference Engine} processes the incoming feature vectors from feature extraction engine. If the inference engine detects an anomaly, it emits an alert containing the metadata (e.g., flow ID, source IP/port, destination IP/port, protocol, etc.). When an alert is received, the \textit{Forensic Module} uses the alert metadata to look up the matching packet traffic from the packet ring buffer and pulls the suspect raw pcap/flow data before it is overwritten. Features extracted from same raw pcap/flow data are forwarded to a global DT-equivalent instance of the NIDS. We use a global explanation method such as TRUSTEE \cite{trustee} to build a Decision tree (DT) that imitates the decisions of the NIDS Inference Model. Relevant windowed packets are obtained from the raw pcap/flow data as explained later. The forensic module packages the windowed packets into an isolated forensic file for the security analysts. Features extracted from windowed packets are forwarded to the \textit{Alert Explanation Module}.

\textbf{Approximating Windowed Packets}. A sliding window of length $k$ goes through the retrieved raw packets and the windowed packets are presented as input to the NIDS-DT instance, which can be described as $f'(X_{t,k})$. To effectively find a small number of inputs to approximate the relevant windowed packets, we find a small $m$ for which ${||f'(X_{t,m}) - f'(X_{t,k})||}_1 < \delta$ and $\delta$ is a small deviation regarding the original detection score. We start with a configurable value for $l$. If ${||f'(X_{t,m}) - f'(X_{t,k})||}_1 \geq \delta$, we update $m = 2m$. If ${||f'(X_{t,m}) - f'(X_{t,k})||}_1 < \delta$, we update $m = \lfloor \frac{m}{2} \rfloor$. There are two termination requirements for this process: the deviation $\delta$, which determines the precise of the approximation; and the number ($U$) of the largest times for updating $m$. We repeat this process until one of the termination requirements is satisfied.

\subsection{Capturing Feature Dependencies During Explanation}
DL-based NIDS usually applies correlation-based methods (e.g., clustering) or domain knowledge to handle feature correlations of structured data for better detection performances.  To capture the feature dependencies of the structured data for an explanation, we first divide the features into several groups. Then, we apply an adaptive sparse group lasso penalization to the feature groups to achieve a sparse explanation. 
\subsubsection{Feature Groups}
\label{feat-grps}
Unlike an image or a block of binary code, where each feature has the same type (e.g., pixel or
hex value), the data samples used for DL-based NIDS are well-structured and have strict formats. For example, Kitsune \cite{kitsune} uses dozens of features from the IP header as input; among them, \textit{TCP.srcport} is a sub-feature of \textit{TCP}, which means if the \textit{TCP.srcport} feature is valid, then the \textit{TCP} feature is supposed to be valid as well. At the same time, the \textit{UDP} is a mutex feature of TCP, which means if the \textit{UDP} feature is valid, then the \textit{TCP} feature is supposed to be invalid. Note that valid means the value of this feature is meaningful. Invalid means the value of this feature is unset. To achieve high-fidelity explanations, we address the feature correlation challenge in structured data by dividing the features of the input into several groups based on their correlations.
Further, some features may belong to multiple groups. For example, the feature $HpHp\_0.1\_weight\_0$ used in Kitsune dataset corresponds to weights aggregated by traffic sent between a set of source and destination IP addresses using a specific protocol (ARP/TCP/UDP) with time window 0.1 (10 seconds). This feature is a sub-feature of ARP, TCP and UDP features. Such features are addressed by using overlapping sums penalty during application of adaptive sparse group lasso to feature groups.
\begin{equation}
    \begin{gathered}
	    x_t^q = \mathds{1}_{A_q} : x_t \\
	    \bigcup_{q=1}^M x_t^q = x_t \; \text{and} \; n(x_t^i \bigcap x_t^j) \geq 0 \; (i \neq j) \\
	    \mathds{1}_{A_q} : x_t \in \{0,1\}, \; j = \left\{ 
	                                           \begin{array}{ll}
	                                           1, & \text{if} \; j \in A_q \\
	                                           0, & \text{otherwise}
	                                           \end{array}
	                                           \right.
	\end{gathered} 
\end{equation} 
where $x_t^q$ is $q^{th}$ group of $x_t$, $M$ is the number of groups within $x_t$, and $\mathds{1}_{A_q} : x_t$ is $q^{th}$ indicator function, which determines whether a feature in $x_t$ belongs to group $x_t^q$. Here each element inside the indicator function is either $0$ or $1$, representing absence or presence, respectively. We need $M$ indicator functions to divide $x_t$ into $M$ groups. A feature may be present in more than one group.

To determine the indicator functions, we consider three scenarios: (1) the grouping strategy used by the target NIDS is available, which is usually determined by the domain knowledge. Then we adopt the same grouping strategy to EXP-SEC; (2) the dataset that contains windowed packet samples are available, but the grouping strategy used by the target NIDS is unclear. Then we calculate the correlations of features with clustering methods to create indicator functions; and (3) the grouping strategy and dataset are unavailable, and EXP-SEC is forced to set the size of each group to one. Consequently, adaptive sparse group lasso will be reduced to lasso. Additionally, we use the same indicator functions for $x_t$ to divide each input $x_i$ inside history inputs ${X'}_{t,m}$ to $M$ groups since every input for NIDS follows the same well-defined structures. We denote the overall number of groups for $x_t$ and ${X'}_{t,m}$ as $Q = M \times (1 + m)$.

\subsubsection{Adaptive Sparse Group Lasso (with Overlapping Groups)} 
To achieve sparse explanations, we need to minimize $\phi(e)$, namely, choose the most relevant features from inputs as explanation results, while omitting those that do not significantly contribute to the detection results. Sparse group lasso (SGL) is a regression method that allows predefined groups of features to be selected into or out of a model together, where all the features of a particular group are either included or excluded. More importantly, it has the desired effect of group-level and feature-wise sparsity [66]. Sparse group lasso allows us to find the important explanatory factors in predicting the corresponding detection result, where each explanatory factor may be comprised of a group of features within the inputs. At the same time, sparse group lasso helps us to achieve the sparse effects on both the group level and the feature level. 

Adaptive sparse group lasso (ASGL) offers more flexibility over sparse group lasso by applying different weights to different groups to adjust the level of penalization. If a group of features is important, it should have a smaller weight and thus be lightly penalized. On the other hand, if a group is not important, it can be assigned a large weight so that it is heavily penalized. Therefore, we use adaptive sparse group lasso to enable xNIDS to meet the sparsity requirements in explaining DL-NIDS. Since some features can belong to multiple groups as illustrated in Section \ref{feat-grps}, we use the overlapping sums penalty in addition (ASGL-O). We model the regression problem as follows:


\begin{multline}
\label{eq-slass-opt}
	\argmin_{\beta} \left\{ {||f-g||}_2^2 + \inf_{\tilde{v_q}\in\mathds{V}_q, \sum_{q=1}^Q\tilde{v_q}=u} (1-\alpha)\lambda\sum_{q=1}^Q \sqrt{p_q}\tilde{v_q}{||\beta_q||}_2 \right. \\
	\left. + \alpha\lambda\sum_{j=1}^p \tilde{w_j}|\beta_j| \right\}
\end{multline}

where $\beta_q$ is a vector, which contains the coefficients for the features in $q^{th}$ group; $p_q$ is the size of $q^{th}$ group; $\beta = (\beta_1 \dots \beta_Q)$; and $\alpha \in [0,1]$ a convex combination of the lasso and group lasso penalties. To achieve group level sparsity, we minimize $\sum{||\beta_q||}_2$, namely exclude more groups by making ${||\beta_q||}_2 = 0$. To achieve feature-level sparsity, we minimize $\sum|\beta_j|$, excluding more features by making $|\beta_j| = 0$. $Q$ is the total number of groups and $\lambda$ is a tuning parameter. 
$\tilde{u_q}\in\mathds{R}^Q$ and $\tilde{w_j}\in\mathds{R}^p$ are known weight vectors which enable adaptive penalization of different feature groups and individual features respectively.

\subsubsection{Model Development}
We integrate the designs discussed above into a consolidated explanation model that explains the results of a DL-based NIDS. First, we use the proposed method in Section \ref{sus-pkt-isol} to approximate the relevant inputs ${X'}_{t,m}$. Finally, following the concept of linear approximation, we use linear components to approximate the local decision boundary of the DL-based NIDS. Specifically, the key idea here is to utilize a local linear model to approximate the individual decision boundary around $y_t = f(X_{t,k})$. The approximation procedure is described as follows:

\begin{equation}
\label{eq-f}
	f(X_{t,k}) = f({X'}_{t,m}) + \delta = g({X'}_{t,m}) + \epsilon	
\end{equation}

where $f(\cdot)$ is the non-linear DL-based NIDS, $g(\cdot)$ is the local approximation method, and $\epsilon$ is a small deviation between the approximation and the true detection result. Guided by feature groups, we translate $g(\cdot)$ into the following equation:

\begin{equation}
\label{eq-g}
	g({X'}_{t,m}) = \sum_{i=t-1}^{t-m} \sum_{q=1}^M x_i^q \beta_{i,q}^T	
\end{equation}

where $x_i^q$ is the $q^{th}$ group from input $x_i$, and $\beta_{i,q}$ contains the corresponding coefficients for $x_i^q$. $\beta_q^T$ contains the coefficients for the features in $q^{th}$ group. $M$ is the number of feature groups within one input. $m$ is the number of history inputs.

Finally, by taking Eq. \ref{eq-f} as the seed input to create $n$ synthesized samples, we formalize our explanation model as the following regression problem:

\begin{multline}
\label{eq-exp-model-opt}
	\argmin_{\beta} \left\{ \frac{1}{2n}{||y-g({Z'}_{t,m})||}_2^2 + (1-\alpha)\lambda\sum_{q=1}^Q \sqrt{p_q}\tilde{v_q}{||\beta_q||}_2 \right. \\ 
	\left. + \alpha\lambda\sum_{j=1}^p \tilde{w_j}|\beta_j| \right\}
\end{multline}

where ${Z'}_{t,m}$ are the synthesised samples for the relevant inputs ${X'}_{t,m}$. $y$ is a vector, containing $n$ detection results. To solve the objective function in Eq. \ref{eq-exp-model-opt}, adaptive sparse group lasso needs to repeat two loops: a group-level outer loop and a feature-level inner loop. The group-level outer loop recurrently checks whether the group’s coefficient is a zero vector. If a group’s coefficients are a non-zero vector, then the feature-level inner-loop revises each parameter within the vector. The adaptive sparse group lasso repeats the two loops until the parameters converge.
The explanation consisting of group-level and feature-level coefficients are passed to the Mapping module (shown in Fig. \ref{exp-sec-arch}) which transforms the prominent features/feature groups into domain-specific explanations for security analysts.

\subsection{Mapping Feature-based Explanation to Domain-Specific Explanation}
In \cite{moustafa-cst}, authors argue that many XAI methods produce low-level explanations, typically in the form of numerical feature importance vectors. These low-level explanations are useful for developers and researchers to understand the internal behaviour of models. However, they lack the contextual interpretation needed for end users (e.g., security analysts). There is a lack of alignment between explanations and the domain-specific knowledge that analysts rely on during investigations. In contrast, high-level explanations aim to relate model outputs to broader security concepts, such as known attack tactics or threat indicators, making them more accessible to security analysts. Such explanations provide the contextual clarity and actionable insights that security analysts need. To address this, \cite{iiot-ids} emphasised the necessity of collaborative efforts in advancing XAI to communicate results effectively to non-AI experts. Furthermore, incorporating domain-specific knowledge into XAI methods is crucial for improving model interpretability and elevating explanations to high-level and actionable insights \cite{xai-scientific}.

\subsubsection{Mapping Module Architecture and Operation}
%

\begin{figure*}[t]
\centering
\includegraphics[scale=0.3]{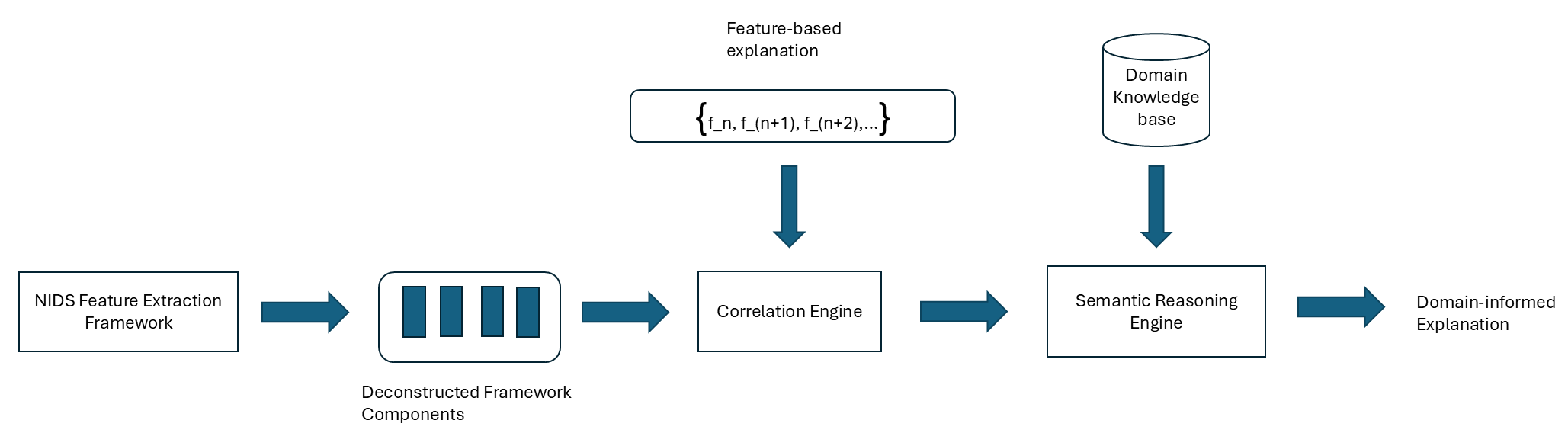}
\caption{Architecture for Mapping Module that translates Feature-based Explanation to Domain-specific Explanation.}
\label{exp-sec-map-mod}
\end{figure*}

The feature/group-level explanation generated by the explanation module of EXP-SEC is passed on to the \textit{Mapping Module}, which aligns those features with domain-specific knowledge, as illustrated in Fig. \ref{exp-sec-map-mod} by: (1) de-constructing the feature extraction framework used by the NIDS model, (2) employing a correlation engine that correlates the explanation features with the de-constructed feature extraction framework's components, and (3) employing a semantic reasoning engine that translates the correlated output of previous stage to offensive tactics derived from domain-specific knowledge bases, expert knowledge, or security frameworks (e.g., Microsoft Cyber Kill Chain, MITRE ATT\&CK framework). The translated tactics outline specific indicators that security analysts can use to identify malicious behaviour. Further, these tactics are enhanced with contextual security framework resources, such as detection recommendations, mitigation strategies, and real world examples.

\subsubsection{Kitsune Feature Mapping Case}
We will take the example of the Kitsune dataset features to describe our domain-specific explanation mapping process. Kitsune NIDS does not read packets similar to a human analyst (e.g., ``This is a malformed HTTP header''). Instead, it looks at the meta-statistics of traffic flows over time. There is a semantic gap between the abstract, mathematical feature vectors used by the model (e.g., ``damped incremental radius of packet size over a 100ms window'') and the tactical domain descriptions required by security analysts to triage threats (e.g., ``ARP Spoofing'', ``Data Exfiltration'', or ``Beaconing''). To bridge this gap for an analyst, we must translate these mathematical observations into behavioral descriptions. By de-constructing the mathematical primitives of the AfterImage feature extraction framework, specifically its implementation of damped incremental statistics, temporal decay windows, and bivariate correlations, we establish a rigorous translation layer. 

\textbf{The Spatial Context (``Who is talking?'')}: Kitsune does not treat all traffic as a single stream. It aggregates traffic into specific streams (or contexts) based on header fields before calculating statistics. The prefix of the feature name tells the analyst what scope of the network is behaving anomalously. Kitsune uses four primary keys plus a specialized jitter key. The names of above feature prefixes, corresponding mapping for a security analyst and security relevance are shown in Table \ref{spatial-map-kitsune-table}. 

\begin{table}[h]
	\centering
    \begin{tabular}{|p{1cm}|p{1cm}|p{2cm}|p{3cm}|}
    \hline
    \textbf{Feature Prefix} & \textbf{Tech Definition} & \textbf{Analyst Mapping} & \textbf{Security Relevance} \\ \hline
    MAC / MI & MAC + IP & Local Device Behavior & This view validates the physical identity of the device. It answers: Is this specific device (laptop, IoT cam) acting strangely on the LAN? (e.g., ARP spoofing, DHCP exhaustion, Lateral movement). \\ \hline
    H & Source IP & Sender Profile & This view aggregates all traffic leaving a specific machine, regardless of destination. It answers: Is this IP acting like an aggressor? (e.g., Scanning (one-to-many), Volumetric Flooding (DoS source), and Botnet infection activity). \\ \hline
    HH & Src IP $\leftrightarrow$ Dst IP & Conversation / Session & This view analyzes the specific conversation between two hosts. It answers: Is the relationship between these two peers normal? (e.g., Lateral movement, C2 communication, Tunneling, Data exfiltration (one-to-one)). \\ \hline
    HpHp & Src IP:Port $\leftrightarrow$ Dst IP:Port & Application / Service & This is the most granular view, analyzing specific process-to-process communication. It answers: Is this specific application protocol behaving as expected? (e.g., Brute force on Port 22, SYN flood on Port 80, Service exploitation, Tunneling, Fuzzing). \\ \hline
    HH\_jit & Jitter & Temporal Cadence of Connection & Focuses on the inter-arrival times of the $\langle SourceIP, DestinationIP \rangle$ channel and measures the rhythm of the connection (e.g., Automated beaconing (regular rhythm) vs. Human activity (irregular rhythm)).
    \end{tabular}
    \caption{Spatial context mapping for Kitsune features}
    \label{spatial-map-kitsune-table}
\end{table}

\textbf{The Statistics (``What are they doing?'')}: For each combination of Time Window and Spatial Context Key, Kitsune calculates a set of statistical features. They are geometric descriptors of the traffic's shape in high-dimensional space. Understanding the meaning of these statistics is crucial for bridging the semantic gap. The statistical features belong to the following categories: (1) \textit{Univariate Statistics}: These statistics apply to a single stream of values (e.g., packet sizes or inter-arrival times), and (2) \textit{Bivariate Statistics}: These statistics describe the relationship between two streams, typically the inbound and outbound traffic of a channel, or the relationship between packet size and count. The names of above statistics, corresponding mapping for a security analyst and security relevance are shown in Table \ref{stat-map-kitsune-table}. 

\begin{table}[h]
	\centering
    \begin{tabular}{|p{1cm}|p{1cm}|p{2cm}|p{3cm}|}
    \hline
    \textbf{Statistics Name} & \textbf{Kitsune Feature} & \textbf{Analyst Mapping} & \textbf{Security Relevance} \\ \hline
    Weight & $w$ & Traffic Intensity / Volume & \textbf{High Value}: "Surge in volume." Indicates Flooding (DoS), Data Exfiltration (large file upload), or aggressive Scanning. \textbf{Low Value}: "Sporadic/Low volume." Indicates stealthy beaconing or "Low-and-Slow" attacks. \\ \hline
    Mean (Average Size) & $\mu$ & Packet Profile & \textbf{Small Mean}: Control-plane traffic (SYNs, ACKs, FINs). A stream of only small packets often indicates a SYN Flood or C2 Heartbeats. \textbf{Large Mean}: Data-plane traffic (Payloads). A stream of consistently large packets often indicates Data Exfiltration or Video Streaming. \\ \hline
    Standard Deviation (Variance) & $\sigma$ & Behavioral Consistency & \textbf{Near Zero (Low Variance)}: The traffic is perfectly uniform (e.g., every packet is exactly 64 bytes). This is a strong indicator of Automated Tools, Bots, or Scripted attacks. \textbf{High Variance}: The traffic size varies wildly. In some contexts (like HTTP), this is normal. In others (like ICMP), it might indicate Tunneling (hiding data in varying packet sizes). \\ \hline
    Magnitude & $||S_i, S_j||$ & Aggregate Channel Throughput & This measures the intensity of the conversation in both directions combined. For example, if an IoT device (e.g., a smart bulb) suddenly exhibits a high magnitude score, it could have been compromised as a bot and is flooding a target. \\ \hline
    Radius & $R$ & Aggregate Channel Volatility & This measures how chaotic the conversation is. For example, if one observes a combination of high magnitude (lots of data) and zero radius (identical packets), it might indicate a volumetric flooding attack. \\ \hline
    Correlation / Covariance & $PCC$ / $Cov$ & Pattern Regularity & \textbf{High Correlation}: Highly organized attack traffic (e.g., a DDoS tool that scales packet size with rate). \textbf{Break in Correlation}: A sudden anomaly where the established pattern of a protocol breaks (e.g., high volume but zero payload, which violates the protocol's normal behavior). \\ \hline
    \end{tabular}
    \caption{Statistics mapping for Kitsune features}
    \label{stat-map-kitsune-table}
\end{table}

\textbf{The Time Windows (``How fast is it happening?'')}: Kitsune utilizes five distinct decay factors (or time windows) for every statistic. They represent distinct temporal resolutions of network behavior. Understanding these windows is the first step in semantic mapping, as the speed of an anomaly is a primary classifier of attack type. 
An \textbf{L1 spike} corresponds to a flash event, likely caused by a script or automated tool. An \textbf{L5 Spike} corresponds to sustained pressure, likely caused by data transfer or a persistent connection state. The names of above time windows, typical time durations, corresponding mapping for a security analyst and security relevance are shown in Table \ref{window-map-kitsune-table}. 

\begin{table}[h]
	\centering
    \begin{tabular}{|p{1cm}|p{1cm}|p{1cm}|p{2cm}|p{3cm}|}
    \hline
    \textbf{Time Window} & \textbf{Decay Factor} ($\lambda$) & \textbf{Typical Duration} & \textbf{Analyst Mapping} & \textbf{Attack Type Indicator} \\ \hline
    L1 & 5 & $\sim$100ms & The Reflex Window (Micro-Burst) & This window captures instantaneous network events. Anomalies here indicate jitter, switching latency issues, or extremely high-velocity attacks like packet spray or rapid-fire port scans. \\ \hline
    L2 & 3 & $\sim$500ms & The Interactive Window (Human Speed) & This timeframe corresponds to human interaction latencies- keystrokes in an SSH session, mouse clicks, or command-line responses. Anomalies here often relate to interactive shell exploits or brute-force login attempts. \\ \hline
    L3 & 1 & $\sim$1.5s - 10s & The Transactional Window (Protocol Cycle) & This captures the duration of typical protocol transactions, such as a full DNS query/response cycle or an HTTP page load. \\ \hline
    L4 & 0.1 & $\sim$10s & The Session Window (Flow Duration) & This window monitors sustained flows, such as buffering a short video clip, a file download, or a series of API calls. It captures the behavior of a session rather than a single packet. \\ \hline
    L5 & 0.01 & $\sim$1 min & The Trend Window (Baseline/Long-Term) & This captures long-term accumulation and background trends. Anomalies here are indicative of ``low-and-slow'' attacks, data exfiltration, or gradual resource exhaustion (e.g., Slowloris) that seek to evade detection by spreading out over time. \\ \hline
    \end{tabular}
    \caption{Time window mapping for Kitsune features}
    \label{window-map-kitsune-table}
\end{table}

If Kitsune triggers an alert corresponding to an attack scenario, the EXP-SEC mapping module uses the de-constructed feature extraction framework as explained above, correlates the raw feature vectors obtained from explanation module with the de-constructed components (spatial, statistics, time window), stitches the correlated information together and plugs the semantic gap to generate a natural-sounding narrative description for the security analyst.

\textbf{Sample Explanations for an Analyst}. Below, we use a few realistic attack scenarios to illustrate the working of the mapping module.\\
\textit{Scenario A: Mirai Botnet Scanning}:
\begin{itemize}
	\item Raw Features: Src\_L3\_weight: HIGH, Src\_L3\_mean: LOW, Src\_L3\_std: LOW
	\item De-constructed and correlated features:
	\begin{itemize}
	    \item Who: The Source IP is generating the anomaly.
		\item What: High volume of traffic (High Weight) consisting of small packets (Low Mean) with identical sizes (Low Std).
		\item How fast: Packets are being generated every few seconds.
    \end{itemize}
	\item Analyst Explanation:
	\begin{itemize}
		\item Detection: ``High-Intensity Robotic Scanning.''
		\item Semantic Context: This matches the profile of an automated botnet scanner (like Mirai) searching for open Telnet ports.
	\end{itemize}
\end{itemize}

\textit{Scenario B: Data Exfiltration}:
\begin{itemize}
	\item Raw Features: Src-Dst\_L5\_weight: MEDIUM, Src-Dst\_L5\_mean: HIGH, Src-Dst\_L5\_std: HIGH
	\item De-constructed and correlated features:
	\begin{itemize}
	    \item Who: A specific conversation between two IPs (Src-Dst).
		\item What: Sustained connection (Long Window) with very large packet sizes (High Mean).
		\item How fast: Packets are being generated every few minutes.
    \end{itemize}
	\item Analyst Explanation:
	\begin{itemize}
		\item Detection: ``Sustained Heavy Data Transfer.''		
		\item Semantic Context: This is a potential data exfiltration attempt or unauthorized file transfer.
	\end{itemize}
\end{itemize}

\textit{Scenario C: ARP Spoofing (Man-in-the-Middle)}:
\begin{itemize}
	\item Raw Features: MI\_L1\_weight: HIGH (on ARP protocol)
	\item De-constructed and correlated features:
	\begin{itemize}
	    \item Who: A MAC address on the local LAN.
		\item What: A sudden, intense burst of traffic (High Weight in Short Window) associated with ARP headers.
		\item How fast: Packets are being generated every few milliseconds.
	\end{itemize}
	\item Analyst Explanation:
	\begin{itemize}
		\item Detection: ``Layer 2 Micro-Burst (ARP Storm).''
		\item Semantic Context: This is likely an ARP Spoofing attempt to hijack traffic.
	\end{itemize}
\end{itemize}

\section{Experimental Results}
\label{results}
\subsection{Implementation and Evaluation Setup}
We evaluate the feature-based explanation component of EXP-SEC in terms of criteria introduced in \cite{sok-dl-sec} as well as new group-level and overlap-aware criteria. In the following experiments, we first scale the importance score vectors $\beta_j$ and $\beta_q$ to the range $[0,1]$. We then rank the features and the groups based on their importance scores. A larger importance score demonstrates a higher relevance of the corresponding feature/group.

\textbf{EXP-SEC Implementation}: While capturing feature dependencies, we follow the group strategies of the target DL-NIDS. We implement EXP-SEC with the Python package \textit{yaglm} \cite{yaglm} which uses an internal grid search to iterate over parameter values for loss functions ($=$ [{`lin\_reg', `l2', `huber', $\dots$]), constraints ($=$ [{`pos', `isotonic', `simplex', $\dots$]) and penalties ($=$ [{`none', `ridge', `lasso', $\dots$]) during cross-validation. We set the number of folds for cross-validation$ = 5$. We tune the parameters on the training dataset and fix them on the testing dataset. 

\textbf{Target NIDSs}: We have used the following state-of-the-art DL-based NIDSs for our analysis: Kitsune \cite{kitsune} with its published dataset; and RNN-IDS \cite{rnn-ids} with the NSL-KDD dataset \cite{nsl-kdd}. For the strategy to divide training and testing data, we follow the original setting in those papers.


\textbf{Comparison Baseline}: We compare EXP-SEC with xNIDS \cite{xnids}, which has been shown to outperform older state-of-the-art explanation methods such as LIME, SHAP, LEMNA, IG, and LRP.

\textbf{Metrics}: We begin by using conventional feature-level metrics such as \textit{descriptive accuracy}, \textit{sparsity} and \textit{stability} \cite{sok-dl-sec} for the performance comparison. We expect EXP-SEC to perform similar to xNIDS as per those metrics since the said metrics are only designed to measure how good an explanation method is at selecting relevant individual features. Then, we proceed to use group-level metrics such as \textit{group-wise deletion AUC} and group overlap-aware metrics such as \textit{group bloat factor}.  


\textbf{Test Environment}: We conduct our experiments on a VMWare ESXi server VM with Intel Xeon Silver 4216 CPU @2.10GHz, 64-bit architecture, 8 cores, 16GB RAM and running Ubuntu 20.04 OS. 
\subsection{Feature-Level Metrics}
\subsubsection{Descriptive Accuracy}
This experiment evaluates how accurately the explanation method captures the important features that contribute to a specific detection result. We adopt the Descriptive Accuracy (DA), which is defined as ${DA}_k(x, f) = f(x|modify(k))$ \cite{sok-dl-sec}. For the anomaly samples, $modify(k)$ nullifies $k$ important features to zero. We expect a significant decrease in Average Descriptive Accuracy (ADA) if the selected important features are relevant to the detection result. As shown in Fig. \ref{ada-plot}, EXP-SEC has a similar ADA decrease as xNIDS for Kitsune NIDS and steeper ADA decrease for RNN-IDS. We then use Table \ref{ada-table} to summarize the area under curve for the ADA curves from Fig. \ref{ada-plot}. 

\begin{figure}[h]
    \centering
    \begin{subfigure}[b]{0.45\textwidth}
        \centering
        \includegraphics[width=\textwidth]{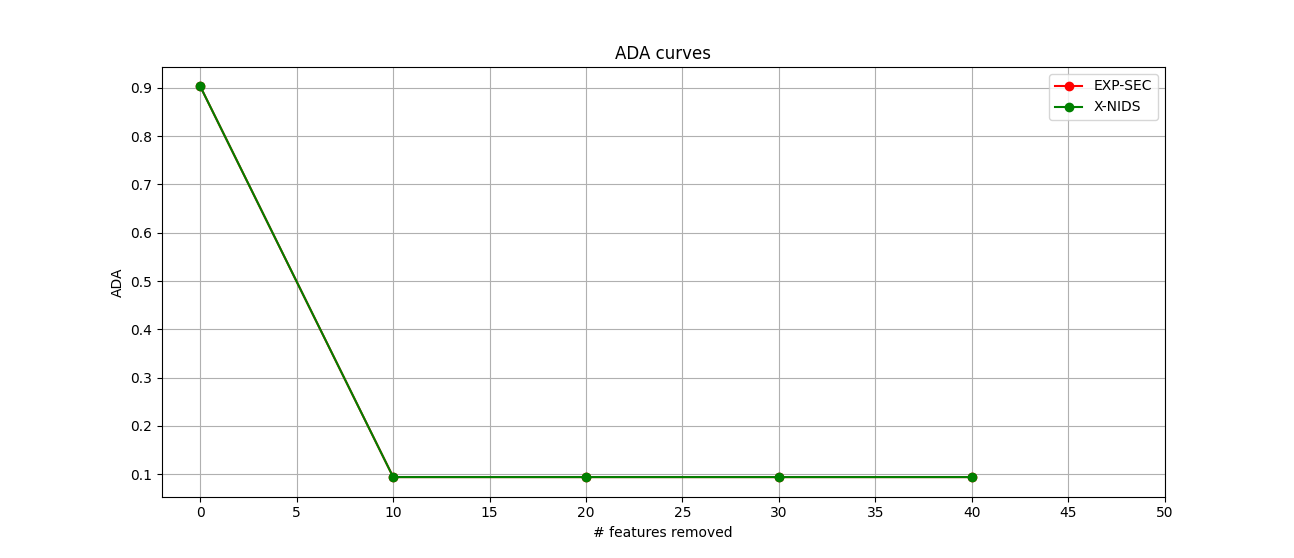}
        \caption{Kitsune}
    \end{subfigure}
    \hfill
    \begin{subfigure}[b]{0.45\textwidth}
        \centering
        \includegraphics[width=\textwidth]{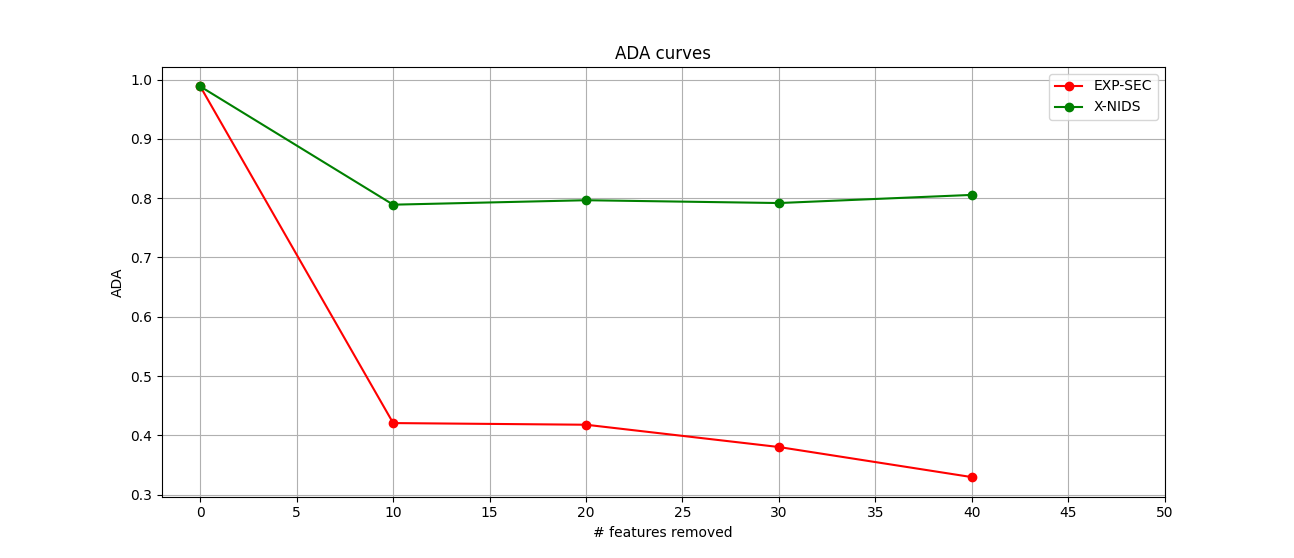}
        \caption{RNN-IDS}
    \end{subfigure}
    \caption{Average Descriptive Accuracy for EXP-SEC and baseline explanation methods.}
    \label{ada-plot}
\end{figure}

\begin{table}[H]
	\centering
    \begin{tabular}{ | l | l | l | }
    \hline
    \textbf{System} & \textbf{Kitsune} & \textbf{RNN-IDS} \\ \hline
    EXP-SEC & 0.19524 & 0.46947 \\ \hline
    xNIDS & 0.19524 & 0.81863 \\ \hline
    \end{tabular}
    \caption{Area under DA curves for explanation methods}
    \label{ada-table}
\end{table}

\subsubsection{Sparsity}
To further evaluate the explanation results, we measure the sparsity of the explanation at the individual feature-level. The desired explanation method should select a limited number of features as explanation results since that would be more convenient for network operators to conduct intrusion analysis and defense. To evaluate sparsity, we follow the Mass Around Zero (MAZ) criteria introduced in \cite{sok-dl-sec}. Since the important scores $\beta = {(\beta_i,\dots, \beta_{\bar{d}})}^T$ are scaled to the range $[0,1]$, we first fit $\beta$ to a half-normalized histogram $h$, we then calculate the MAZ by $MAZ(\beta) = \int_0^1 h(x)dx$ for $\beta \in [0,1]$.

We present the MAZ curves in Fig. \ref{maz-plot}. We also calculate the Area Under Curve (AUC) and show the information in Table \ref{maz-table}. We expect a steep slope near zero from the MAZ curve and a large AUC, if the explanation method can assign zeros to most of the features and achieve sparse explanations. As shown in Fig \ref{maz-plot}, we observe that the steepness of slopes for EXP-SEC and xNIDS are similar, as is the AUC shown in Table \ref{maz-table}, which confirms that EXP-SEC performs similar to xNIDS regarding the sparsity criteria.

\begin{figure}[h]
    \centering
    \begin{subfigure}[b]{0.45\textwidth}
        \centering
        \includegraphics[width=\textwidth]{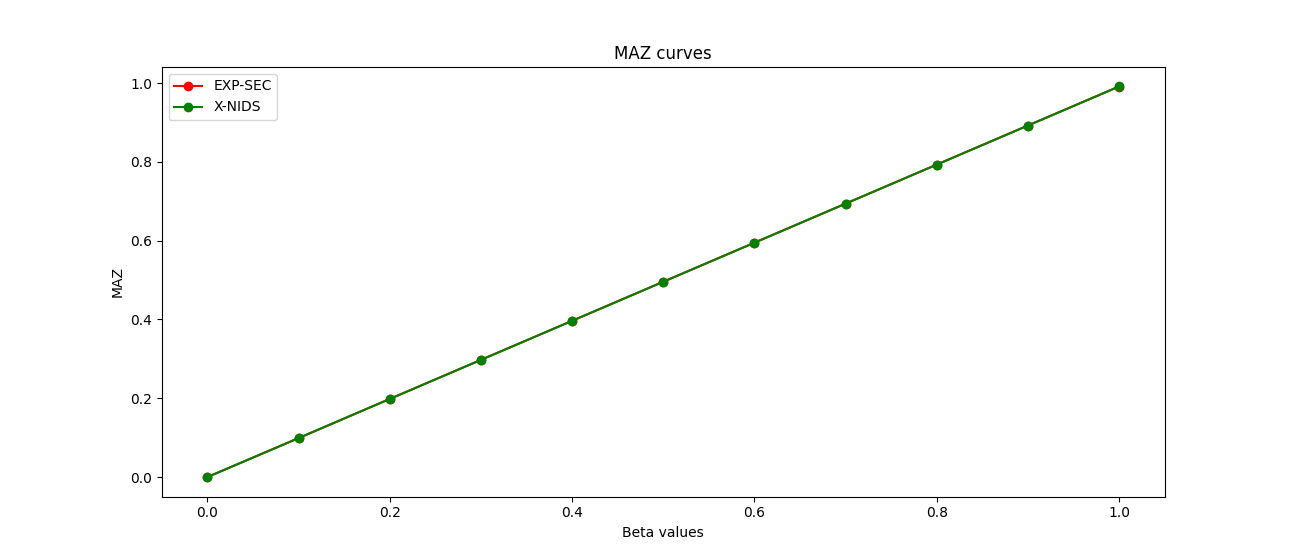}
        \caption{Kitsune}
    \end{subfigure}
    \hfill
    \begin{subfigure}[b]{0.45\textwidth}
        \centering
        \includegraphics[width=\textwidth]{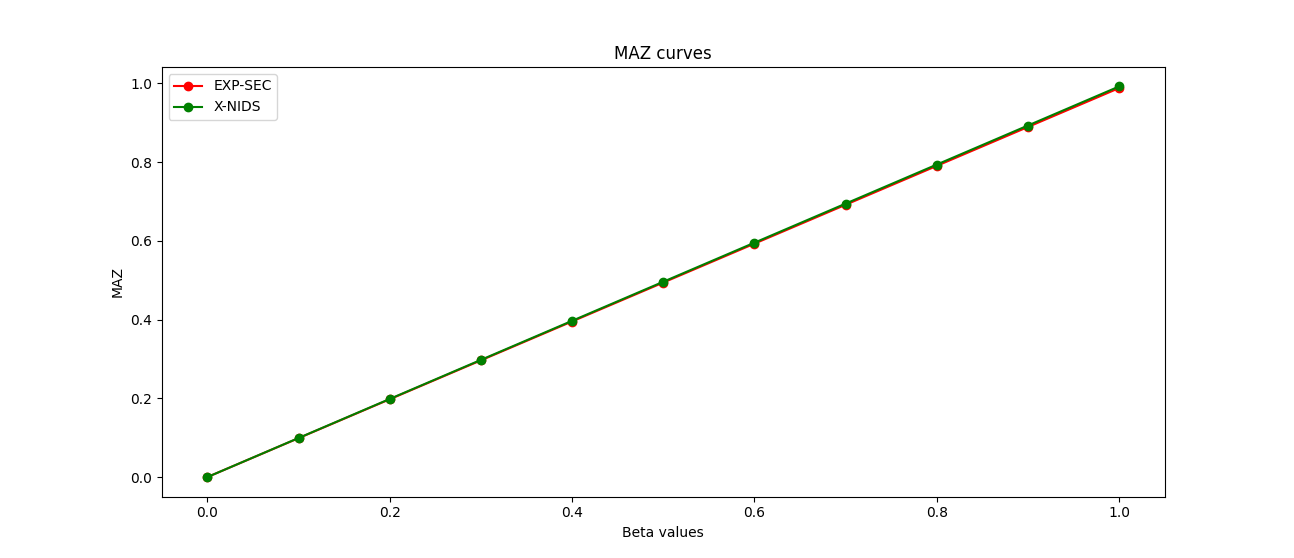}
        \caption{RNN-IDS}
    \end{subfigure}
    \caption{Sparsity for EXP-SEC and baseline explanation methods.}
    \label{maz-plot}
\end{figure}

\begin{table}[H]
	\centering
    \begin{tabular}{ | l | l | l | }
    \hline
    \textbf{System} & \textbf{Kitsune} & \textbf{RNN-IDS} \\ \hline
    EXP-SEC & 0.4956 & 0.49385 \\ \hline
    xNIDS & 0.49568 & 0.49593 \\ \hline
    \end{tabular}
    \caption{Area under MAZ curves for explanation methods}
    \label{maz-table}
\end{table}

\subsubsection{Stability}
We further evaluate the stability of the explanation results following the definition in \cite{sok-dl-sec}. From a network operator’s perspective, the explanation results should be stable. Hence, the desired explanation method should generate similar results for the same samples among multiple tests. To examine the stability of explanations, we calculate the intersection size of the top $K$ features of the explanation results for the same input regarding different tests. We denote the explanation result of the first test as $\beta^1 = {(\beta^1_i ,\dots, \beta^1_{\bar{d}})}^T$ , and for $n^{th}$ test as $\beta^n = {(\beta^n_i ,\dots, \beta^n_{\bar{d}})}^T$. $top(\beta^n)$ denotes the top $K$ features of $\beta^n$. Then the stability is measured as $stability = \frac{1}{K} {||{top(\beta^1) \cap top(\beta^2)\dots \cap top(\beta^n)}||}_1$. A stable explanation method should have a stability score close to $1$. From Table \ref{stab-table}, we can observe that EXP-SEC has slightly lower stability scores than xNIDS. The stability of EXP-SEC explanations depends on the stability of underlying ASGL-O implementation. By controlling the various sources of randomness in the implementation (e.g., seed to adaptive initializer's internal cross-validation fit which produces stage-1 coefficients), EXP-SEC stability can be improved further. 

\begin{table}[H]
	\centering
    \begin{tabular}{ | l | l | l | }
    \hline
    \textbf{System} & \textbf{Kitsune} & \textbf{RNN-IDS} \\ \hline
    EXP-SEC & 0.85 & 0.42 \\ \hline 
    xNIDS & 0.9 & 0.5 \\ \hline
    \end{tabular}
    \caption{Average \textit{stability} scores for explanation methods}
    \label{stab-table}
\end{table}

\subsection{Group-level Metrics}
\subsubsection{Group-wise Deletion AUC}
We rank feature groups by their structural importance score (derived from the $L_2$ norm of the latent groups in ASGL-O) and then progressively mask/remove the top-ranked groups from the input and feed the modified sample back into the NIDS model, tracking the decline in the model's output confidence. We plot the output confidence and measure the Area Under the Curve (AUC). A lower Deletion AUC indicates that the explanation successfully identified the most critical groups first. Standard SGL (xNIDS) suffers from uniform shrinkage bias (over-penalizing the strongest signals) and group fragmentation. As a result, it frequently mis-ranks the true causal features. ASGL-O (EXP-SEC) isolates the exact intersecting groups driving the NIDS model, producing a sharper, faster drop in prediction confidence and a lower Deletion AUC as is confirmed from Fig. \ref{grp-ada-plot} and Table \ref{dauc-table} respectively.

\begin{figure}[h]
    \centering
    \begin{subfigure}[b]{0.45\textwidth}
        \centering
        \includegraphics[width=\textwidth]{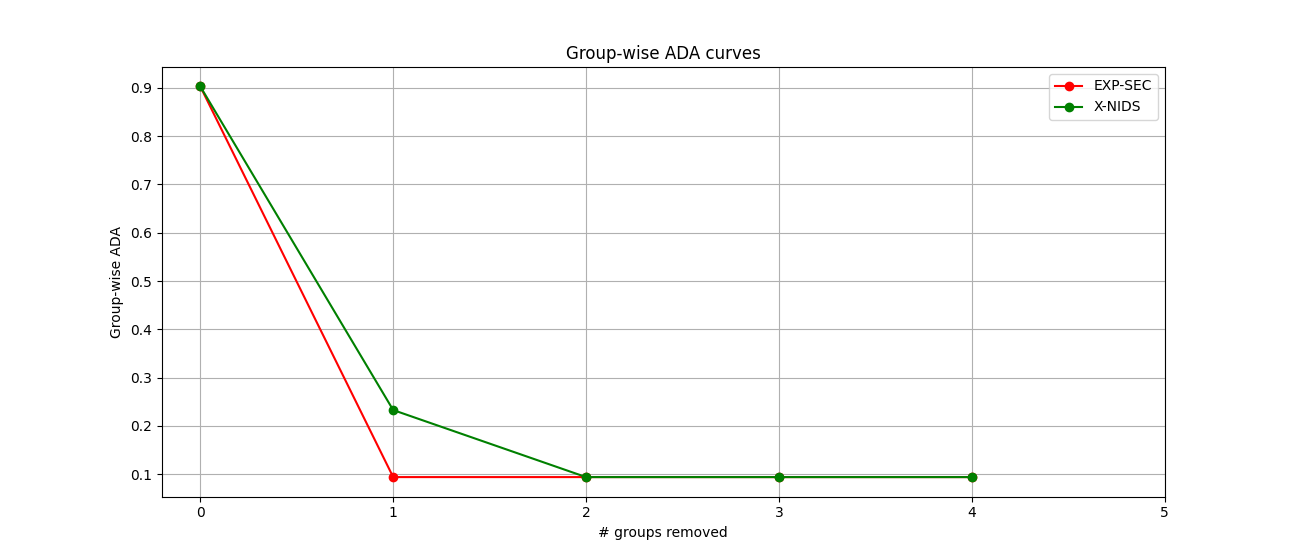}
        \caption{Kitsune}
    \end{subfigure}
    \hfill
    \begin{subfigure}[b]{0.45\textwidth}
        \centering
        \includegraphics[width=\textwidth]{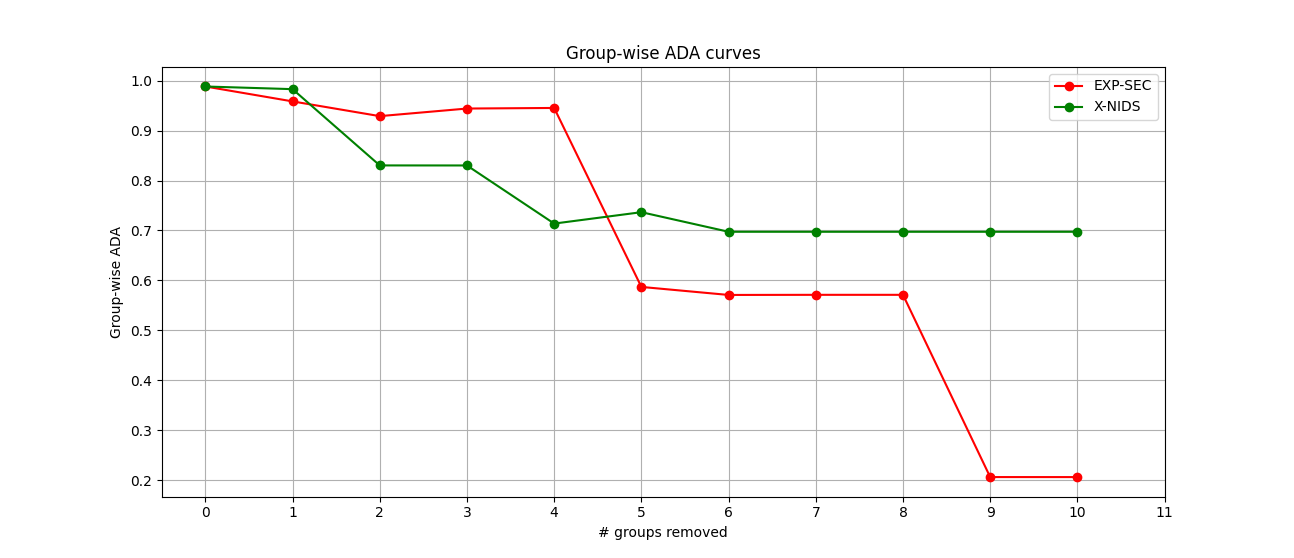}
        \caption{RNN-IDS}
    \end{subfigure}
    \caption{Average Group-wise Prediction Confidence for EXP-SEC and baseline explanation methods.}
    \label{grp-ada-plot}
\end{figure}

\begin{table}[H]
	\centering
    \begin{tabular}{ | l | l | l | }
    \hline
    \textbf{System} & \textbf{Kitsune} & \textbf{RNN-IDS} \\ \hline
    EXP-SEC & 0.19524 & 0.688057 \\ \hline
    xNIDS & 0.22994 & 0.77267 \\ \hline
    \end{tabular}
    \caption{Deletion Area under curve values for explanation methods}
    \label{dauc-table}
\end{table}

\subsection{Overlap-aware Metrics}
\subsubsection{Group Bloat Factor}
Standard SGL (xNIDS) assumes feature groups are completely mutually exclusive (a strict partition) and applies a uniform penalty. ASGL-O (EXP-SEC) introduces both data-dependent weights (to eliminate bias) and latent variable duplication (to natively handle overlapping structural relationships). When used as local surrogates to explain a black-box deep learning model, standard SGL will severely misrepresent the model's decision boundaries wherever features interact across multiple contexts (like a packet belonging to a specific network flow and a specific time frame simultaneously). The Group Bloat Factor (GBF) metric assesses whether the explanation accurately captures the true underlying group structure of the data without generating fragmented results. When a deep learning model triggers an alert based on a feature that inherently lives in multiple groups, standard SGL gets confused. To evaluate this, we measure the total number of active groups returned by the surrogate:
\begin{equation}
    \textsf{GBF} = \frac{|\mathcal{G}_{\text{active}}|}{|\mathcal{G}_{\text{true causal}}|}
\end{equation}

For RNN-IDS, standard SGL exhibits higher value of average GBF ($1.33$) than ASGL-O ($1.00$). Because SGL cannot split a feature's identity across overlapping contexts, it is forced to activate the union of all independent partitions containing that feature. ASGL-O utilizes latent variable space to activate only the precise overlapping group paths that matter, keeping the explanation clean, concise, and highly localized. Since Kitsune dataset features do not exhibit overlap between groups, we excluded it from the GBF evaluation. 

\subsection{Time Efficiency}
We finally examine the time efficiency of EXP-SEC and baseline explanation methods. Table \ref{runtime-table} shows the average run-time per input for all explanations methods and NIDSs. We observe that EXP-SEC is on average $\sim 10^3$ times slower than xNIDS. This is because ASGL-O used by EXP-SEC is much more computationally expensive than SGL used by xNIDS. ASGL-O is a two-step estimator which requires running an initial unpenalized or ridge-penalized regression to calculate the adaptive weights before it can execute the actual sparse group lasso optimization.
Further, if a packet dataset has $P$ features and highly overlapping groups, the latent space expands to $P' \gg P$ features. ASGL-O requires running an initial model in this expanded space just to get the adaptive weights, followed by the actual overlapping group lasso optimization pass. SGL computes its penalties in a single pass. EXP-SEC is, therefore, more suited as an offline explanation framework.


\begin{table}[H]
	\centering
    \begin{tabular}{ | l | l | l | }
    \hline
    \textbf{System} & \textbf{Kitsune} & \textbf{RNN-IDS} \\ \hline
    EXP-SEC & 25.3401 & 23.3089 \\ \hline 
    xNIDS & 0.021735 & 0.044285 \\ \hline
    \end{tabular}
    \caption{Runtime per sample (in seconds) for explanation methods}
    \label{runtime-table}
\end{table}

\section{Discussion}
\label{discuss}

An real-world explanation framework designed for NIDS deployed in enterprise networks must be highly adaptable. Network traffic is dynamic and a rigid explanation framework will quickly suffer from performance bottlenecks, generate stale or inaccurate interpretations or become a vulnerability itself.
\textbf{Traffic Volatility}. Network throughput can fluctuate significantly, e.g., standard business hours vs. a sudden 100Gbps DDoS attack. The explanation framework must dynamically scale its computational complexity based on the network's current load. Under normal traffic conditions, the framework can use maximum fidelity by generating a high number of perturbations for an ASGL-O surrogate to give pinpoint, packet-level accuracy. During an attack spike or high-load event, the framework must automatically pivot to a low-latency mode (e.g., reducing the number of local perturbation samples, or aggregating features at the flow level rather than the packet level) to avoid exhausting system memory and causing packet drops. 

\textbf{Changing Network Topologies}. Enterprise networks are constantly evolving. New subnets can be provisioned, microservices can scale up or down, and protocols can shift (e.g., transition from HTTP/2 to HTTP/3). The structural groups used by the lasso surrogate (flows, time-windows, protocol families) cannot be hardcoded into the architecture. They must be determined dynamically. The Stateful Stream Engine must ingest real-time network configuration data (such as asset inventories or routing tables). If a new cluster of IoT devices is deployed, the system must automatically create a new behavioral overlapping group category for those devices, allowing the explainer to instantly evaluate them without requiring a manual code re-factor.

\textbf{NIDS Model Upgrades}. Security teams frequently update, retrain, or completely swap out their NIDS engines. The core optimization layer of the explainer must decoupled from the NIDS model's internal architecture. Relying on a local surrogate model (like ASGL-O) provides natural agnosticism because it only requires black-box API access (inputs and output probabilities). However, the system should feature a hybrid execution layer: if the NIDS is a black-box, it defaults to surrogate sampling; if the NIDS is upgraded to a differentiable deep learning model, the framework should adaptively unlock faster, white-box gradient attribution methods (like Integrated Gradients) to save compute.

\textbf{Adversarial Evasion}. Sophisticated attackers may figure out how an XAI framework perturbs data to generate explanations. Based on that information, they can craft malicious traffic wrapped in an adversarial layer. This wrapper detects when it is being probed by the XAI system and temporarily forces the NIDS to output a benign score, masking the attack. The explanation framework’s perturbation and sampling engine must be dynamic and unpredictable to an outside observer. The framework should adaptively alter its perturbation strategies by changing the distribution of its masking noise, mixing up its time-window lengths, or utilizing context-aware generative models (like GANs or autoencoders trained on normal traffic) to ensure that perturbed packets look indistinguishable from real, protocol-compliant network traffic.

\section{Related Work}
\label{literature}
\textbf{Why explanation for ML-based security systems is needed?} Several works have recommended employing explanation techniques for ML systems deployed for cybersecurity. In \cite{dos-n-donts-ml}, the authors have recommended employing explanation techniques to delve deeper into the features of learning-based security systems. Not withstanding their limitations, these techniques can unveil spurious correlations and empower experts to evaluate their effect on the security system's features. The authors in \cite{ml-insec} have also emphasized the importance of explaining the outcomes produced by ML-based detectors used in security contexts. Beyond mere predictions, a cybersecurity threat detection model can offer valuable insights. By understanding the model’s explanations, security practitioners can enhance protection around monitored assets in subsequent executions. Explainability varies in relevance across different domains, but the authors contend that it is crucial for several cybersecurity tasks, enabling the effective application of countermeasures against security threats. The participants (security practitioners) of a survey conducted by the authors of \cite{pragmatic-ml} have expressed the view that security solution providers ought to prioritize methods that offer clear explanations to their clients. 

\textbf{Application of XAI methods to ML-based security systems}. As a result of above, we can see the beginning of efforts in the research community to apply XAI techniques to ML-based security systems. A few works \cite{kalakoti, patil, arreche, barnard} have employed XAI methods such as LIME \cite{lime} and SHAP \cite{shap} to assess the quality of local/global explanations generated for conventional machine learning classifiers (AdaBoost, k-nearest neighbour, Multi-layer Perceptron, Random Forest, XGBoost, Light Gradient-Boosting Machine, Gradient Boosting Classifier) trained on network intrusion datasets (NSL-KDD dataset \cite{nsl-kdd}, CICIDS-2017 \cite{cicids-2017}) in terms of criteria such as faithfulness, stability, complexity, and sensitivity. In \cite{warnecke}, the authors have delved into six explanation methods, evaluating their effectiveness across four security systems described in existing literature. These systems leverage DL techniques to detect Android malware, malicious PDF files and security vulnerabilities. The assessment criteria encompass both general properties of deep learning and domain-specific aspects relevant to security.

The authors in \cite{maonan} have introduced a framework rooted in SHAP to provide explanations for IDSs. This framework blends both local and global explanations, enhancing the overall interpretability of IDSs. Local explanations shed light on why the ML model employed by a specific IDS reaches particular decisions for individual inputs. Meanwhile, global explanations highlight crucial features extracted from the ML model and elucidate the relationships between feature values and specific attack types. \cite{shtayat} has presented a deep learning-based IDS for IIoT networks consisting of an ensemble of three CNN models and an extreme-learning machine model. The authors have used SHAP and LIME techniques to explain the ensemble detector’s decision-making process. However, most of above works apply existing XAI techniques to security systems including IDS. They do not present original explanation methods targeted at explaining DL-based NIDS.

\textbf{ML-based security systems with built-in explainability}. A few works have also proposed ML/DL-based attack detection systems with built-in explainability. In \cite{doh-explain}, the authors have implemented an XAI solution to provide accurate detection and classification of the DNS-over-HTTPS attacks. It is based on a balanced stacked random forest classifier. The authors have highlighted the underlying feature contributions to provide transparent and explainable results from the model. In \cite{ensemble-explain}, the authors have proposed an NIDS for IT networks combining ensemble learning and stacking with a meta-learner (CNN) that works on graphical representation of traffic flows and provides the required explainability level for the decisions made. They also provide visual representations of network anomalies that allows security analysts to interpret and gain insights into the detected network anomalies. However, most of these systems are targeted at specific attacks (e.g., DNS-over-HTTPS attacks). 

\textbf{Adversarial learning in NIDS}. Though not targeted at generating explanations, recent works on adversarial learning in NIDSs aim to generate specific examples/feature vectors which can evade detection. Based on a complete/partial knowledge of the ML model underlying a given NIDS and the features used to the train the model, researchers have used existing techniques such as genetic algorithm/particle swarm optimization/generative adversarial networks to generate adversarial examples \cite{bastian-1}. An update to that work focuses on restricting the adversarial feature space so that the generated features correspond to functional network packets \cite{bastian-2}. It does that by presenting a constrained optimization formulation for perturbing raw packet payloads while avoiding modification to the original packet function. A meta-heuristic inspired by genetic algorithm is proposed to solve the optimization problem. However, these works do not offer an explanation as to why the adversarial examples generated cause mis-classification by the underlying ML model.

\textbf{Explanation methods for ML-based security systems}. CADE \cite{cade} introduces a distance-based explanation method for security applications. Unfortunately, while it works well for malware detection, it suffers from an extremely low fidelity when applying to DL-based NIDS. DeepAID \cite{deepaid} is a whitebox explanation method based on back-propagation, which focuses on neural network components of DL-based NIDS and ignores both the feature extractor and the feature mapper of Kitsune. Moreover, by assuming each feature is independent, DeepAID is insufficient to capture the feature dependencies of structured data when explaining DL-based NIDS.

\section{Conclusion}
\label{conclusion}
In this paper, we present \textsf{EXP-SEC}, a novel framework which can generate explanations for intrusion detection decisions of DL-based NIDS and is targeted at enterprise SOC analysts. EXP-SEC can isolate the suspect packets/flow which likely caused an alert, handle much more complex feature dependencies in network traffic than existing methods and translate the feature/group-based explanations to domain-specific explanations suitable for processing by security analysts. The evaluation results show that EXP-SEC outperforms xNIDS (existing best performing explanation framework) in terms of group-level and overlap-aware explanation utility metrics while performing similarly in terms of conventional feature-level metrics such as descriptive accuracy, sparsity and stability. Additionally, we demonstrate the security analyst-friendly explanation format generated by EXP-SEC taking the case of Kitsune NIDS.


\bibliographystyle{elsarticle-num}
\bibliography{XAIbib}

\newcommand{\noop}[1]{}
\begin{thebibliography}{10}
\expandafter\ifx\csname url\endcsname\relax
  \def\url#1{\texttt{#1}}\fi
\expandafter\ifx\csname urlprefix\endcsname\relax\def\urlprefix{URL }\fi
\expandafter\ifx\csname href\endcsname\relax
  \def\href#1#2{#2} \def\path#1{#1}\fi

\bibitem{xnids}
F.~Wei, H.~Li, Z.~Zhao, H.~Hu,
  \href{https://www.usenix.org/conference/usenixsecurity23/presentation/wei-feng}{{xNIDS}:
  Explaining deep learning-based network intrusion detection systems for active
  intrusion responses}, in: 32nd USENIX Security Symposium (USENIX Security
  23), USENIX Association, Anaheim, CA, 2023, pp. 4337--4354.
\newline\urlprefix\url{https://www.usenix.org/conference/usenixsecurity23/presentation/wei-feng}

\bibitem{bro}
V.~Paxson, {Bro: a system for detecting network intruders in real-time}, in:
  Proceedings of the 7th Conference on USENIX Security Symposium - Volume 7,
  SSYM'98, USENIX Association, USA, 1998, p.~3.

\bibitem{snort}
M.~Roesch, {Snort - Lightweight Intrusion Detection for Networks}, in:
  Proceedings of the 13th USENIX Conference on System Administration, LISA '99,
  USENIX Association, USA, 1999, p. 229–238.

\bibitem{dd-nids}
D.~Chou, M.~Jiang, \href{https://doi.org/10.1145/3472753}{{A Survey on
  Data-driven Network Intrusion Detection}}, ACM Comput. Surv. 54~(9) (oct
  2021).
\newblock \href {https://doi.org/10.1145/3472753} {\path{doi:10.1145/3472753}}.
\newline\urlprefix\url{https://doi.org/10.1145/3472753}

\bibitem{MLIDissues}
R.~Sommer, V.~Paxson, {Outside the Closed World: On Using Machine Learning for
  Network Intrusion Detection}, in: 2010 IEEE Symposium on Security and
  Privacy, 2010, pp. 305--316.
\newblock \href {https://doi.org/10.1109/SP.2010.25}
  {\path{doi:10.1109/SP.2010.25}}.

\bibitem{kitsune}
Y.~Mirsky, T.~Doitshman, Y.~Elovici, A.~Shabtai,
  \href{http://wp.internetsociety.org/ndss/wp-content/uploads/sites/25/2018/02/ndss2018\_03A-3\_Mirsky\_paper.pdf}{{Kitsune:
  An Ensemble of Autoencoders for Online Network Intrusion Detection}}, in:
  25th Annual Network and Distributed System Security Symposium, {NDSS} 2018,
  San Diego, California, USA, February 18-21, 2018, The Internet Society, 2018.
\newline\urlprefix\url{http://wp.internetsociety.org/ndss/wp-content/uploads/sites/25/2018/02/ndss2018\_03A-3\_Mirsky\_paper.pdf}

\bibitem{dos-n-donts-ml}
D.~Arp, E.~Quiring, F.~Pendlebury, A.~Warnecke, F.~Pierazzi, C.~Wressnegger,
  L.~Cavallaro, K.~Rieck,
  \href{https://www.usenix.org/conference/usenixsecurity22/presentation/arp}{{Dos
  and Don{\textquoteright}ts of Machine Learning in Computer Security}}, in:
  31st USENIX Security Symposium (USENIX Security 22), USENIX Association,
  Boston, MA, 2022, pp. 3971--3988.
\newline\urlprefix\url{https://www.usenix.org/conference/usenixsecurity22/presentation/arp}

\bibitem{ayush-iot-nids-explain}
A.~Kumar, V.~L.~L. Thing, \href{https://arxiv.org/abs/2408.14040}{{Evaluating
  The Explainability of State-of-the-Art Deep Learning-based Network Intrusion
  Detection Systems}} (2024).
\newblock \href {http://arxiv.org/abs/2408.14040} {\path{arXiv:2408.14040}}.
\newline\urlprefix\url{https://arxiv.org/abs/2408.14040}

\bibitem{warnecke}
A.~Warnecke, D.~Arp, C.~Wressnegger, K.~Rieck,
  \href{https://doi.ieeecomputersociety.org/10.1109/EuroSP48549.2020.00018}{{Evaluating
  Explanation Methods for Deep Learning in Security}}, in: 2020 IEEE European
  Symposium on Security and Privacy (EuroS\&P), IEEE Computer Society, Los
  Alamitos, CA, USA, 2020, pp. 158--174.
\newblock \href {https://doi.org/10.1109/EuroSP48549.2020.00018}
  {\path{doi:10.1109/EuroSP48549.2020.00018}}.
\newline\urlprefix\url{https://doi.ieeecomputersociety.org/10.1109/EuroSP48549.2020.00018}

\bibitem{maonan}
M.~Wang, K.~Zheng, Y.~Yang, X.~Wang, {An Explainable Machine Learning Framework
  for Intrusion Detection Systems}, IEEE Access 8 (2020) 73127--73141.
\newblock \href {https://doi.org/10.1109/ACCESS.2020.2988359}
  {\path{doi:10.1109/ACCESS.2020.2988359}}.

\bibitem{cade}
L.~Yang, W.~Guo, Q.~Hao, A.~Ciptadi, A.~Ahmadzadeh, X.~Xing, G.~Wang,
  \href{https://www.usenix.org/conference/usenixsecurity21/presentation/yang-limin}{{CADE:
  Detecting and Explaining Concept Drift Samples for Security Applications}},
  in: 30th USENIX Security Symposium (USENIX Security 21), USENIX Association,
  2021, pp. 2327--2344.
\newline\urlprefix\url{https://www.usenix.org/conference/usenixsecurity21/presentation/yang-limin}

\bibitem{deepaid}
D.~Han, Z.~Wang, W.~Chen, Y.~Zhong, S.~Wang, H.~Zhang, J.~Yang, X.~Shi, X.~Yin,
  \href{https://doi.org/10.1145/3460120.3484589}{{DeepAID: Interpreting and
  Improving Deep Learning-based Anomaly Detection in Security Applications}},
  in: Proceedings of the 2021 ACM SIGSAC Conference on Computer and
  Communications Security, CCS '21, Association for Computing Machinery, New
  York, NY, USA, 2021, p. 3197–3217.
\newblock \href {https://doi.org/10.1145/3460120.3484589}
  {\path{doi:10.1145/3460120.3484589}}.
\newline\urlprefix\url{https://doi.org/10.1145/3460120.3484589}

\bibitem{grad}
K.~Simonyan, A.~Vedaldi, A.~Zisserman, {Deep Inside Convolutional Networks:
  Visualising Image Classification Models and Saliency Maps}, in: Workshop at
  International Conference on Learning Representations, 2014.

\bibitem{ig}
M.~Sundararajan, A.~Taly, Q.~Yan, {Axiomatic attribution for deep networks},
  in: Proceedings of the 34th International Conference on Machine Learning -
  Volume 70, ICML'17, JMLR.org, 2017, p. 3319–3328.

\bibitem{lrp}
S.~Bach, A.~Binder, G.~Montavon, F.~Klauschen, K.-R. M{\"u}ller, W.~Samek,
  \href{https://api.semanticscholar.org/CorpusID:9327892}{{On Pixel-Wise
  Explanations for Non-Linear Classifier Decisions by Layer-Wise Relevance
  Propagation}}, PLoS ONE 10 (2015).
\newline\urlprefix\url{https://api.semanticscholar.org/CorpusID:9327892}

\bibitem{lime}
M.~T. Ribeiro, S.~Singh, C.~Guestrin,
  \href{https://doi.org/10.1145/2939672.2939778}{{"Why Should I Trust You?":
  Explaining the Predictions of Any Classifier}}, KDD '16, Association for
  Computing Machinery, New York, NY, USA, 2016.
\newblock \href {https://doi.org/10.1145/2939672.2939778}
  {\path{doi:10.1145/2939672.2939778}}.
\newline\urlprefix\url{https://doi.org/10.1145/2939672.2939778}

\bibitem{shap}
S.~M. Lundberg, S.-I. Lee, {A unified approach to interpreting model
  predictions}, in: Proceedings of the 31st International Conference on Neural
  Information Processing Systems, NIPS'17, Curran Associates Inc., Red Hook,
  NY, USA, 2017, p. 4768–4777.

\bibitem{lemna}
W.~Guo, D.~Mu, J.~Xu, P.~Su, G.~Wang, X.~Xing,
  \href{https://doi.org/10.1145/3243734.3243792}{{LEMNA: Explaining Deep
  Learning based Security Applications}}, in: Proceedings of the 2018 ACM
  SIGSAC Conference on Computer and Communications Security, CCS '18,
  Association for Computing Machinery, New York, NY, USA, 2018, p. 364–379.
\newblock \href {https://doi.org/10.1145/3243734.3243792}
  {\path{doi:10.1145/3243734.3243792}}.
\newline\urlprefix\url{https://doi.org/10.1145/3243734.3243792}

\bibitem{yaglm}
I.~Carmichael, T.~Keefe, N.~Giertych, J.~P. Williams,
  \href{https://arxiv.org/abs/2110.05567}{{yaglm: a Python package for fitting
  and tuning generalized linear models that supports structured, adaptive and
  non-convex penalties}} (2021).
\newblock \href {http://arxiv.org/abs/2110.05567} {\path{arXiv:2110.05567}}.
\newline\urlprefix\url{https://arxiv.org/abs/2110.05567}

\bibitem{rnn-ids}
C.~Yin, Y.~Zhu, J.~Fei, X.~He, {A Deep Learning Approach for Intrusion
  Detection Using Recurrent Neural Networks}, IEEE Access 5 (2017)
  21954--21961.
\newblock \href {https://doi.org/10.1109/ACCESS.2017.2762418}
  {\path{doi:10.1109/ACCESS.2017.2762418}}.

\bibitem{shapley-val}
L.~S. Shapley, \href{https://doi.org/10.1515/9781400881970-018}{{17. A Value
  for n-Person Games}}, Princeton University Press, Princeton, 1953, pp.
  307--318 [cited 2024-07-17].
\newblock \href {https://doi.org/doi:10.1515/9781400881970-018}
  {\path{doi:doi:10.1515/9781400881970-018}}.
\newline\urlprefix\url{https://doi.org/10.1515/9781400881970-018}

\bibitem{trustee}
A.~S. Jacobs, R.~Beltiukov, W.~Willinger, R.~A. Ferreira, A.~Gupta, L.~Z.
  Granville, \href{https://doi.org/10.1145/3548606.3560609}{{AI/ML for Network
  Security: The Emperor has no Clothes}}, in: Proceedings of the 2022 ACM
  SIGSAC Conference on Computer and Communications Security, CCS '22,
  Association for Computing Machinery, New York, NY, USA, 2022, p. 1537–1551.
\newblock \href {https://doi.org/10.1145/3548606.3560609}
  {\path{doi:10.1145/3548606.3560609}}.
\newline\urlprefix\url{https://doi.org/10.1145/3548606.3560609}

\bibitem{moustafa-cst}
N.~Moustafa, N.~Koroniotis, M.~Keshk, A.~Y. Zomaya, Z.~Tari, {Explainable
  Intrusion Detection for Cyber Defences in the Internet of Things:
  Opportunities and Solutions}, IEEE Communications Surveys \& Tutorials 25~(3)
  (2023) 1775--1807.
\newblock \href {https://doi.org/10.1109/COMST.2023.3280465}
  {\path{doi:10.1109/COMST.2023.3280465}}.

\bibitem{iiot-ids}
M.~M. Shtayat, M.~K. Hasan, R.~Sulaiman, S.~Islam, A.~U.~R. Khan, {An
  Explainable Ensemble Deep Learning Approach for Intrusion Detection in
  Industrial Internet of Things}, IEEE Access 11 (2023) 115047--115061.
\newblock \href {https://doi.org/10.1109/ACCESS.2023.3323573}
  {\path{doi:10.1109/ACCESS.2023.3323573}}.

\bibitem{xai-scientific}
Z.~Wu, J.~Chen, Y.~Li, Y.~Deng, H.~Zhao, C.-Y. Hsieh, T.~Hou, {From Black Boxes
  to Actionable Insights: A Perspective on Explainable Artificial Intelligence
  for Scientific Discovery}, Journal of Chemical Information and Modeling
  63~(24) (2023) 7617--7627, pMID: 38079566.
\newblock \href {https://doi.org/10.1021/acs.jcim.3c01642}
  {\path{doi:10.1021/acs.jcim.3c01642}}.

\bibitem{sok-dl-sec}
A.~Warnecke, D.~Arp, C.~Wressnegger, K.~Rieck, Evaluating explanation methods
  for deep learning in security, in: 2020 IEEE European Symposium on Security
  and Privacy (EuroS\&P), 2020, pp. 158--174.
\newblock \href {https://doi.org/10.1109/EuroSP48549.2020.00018}
  {\path{doi:10.1109/EuroSP48549.2020.00018}}.

\bibitem{nsl-kdd}
B.~Ingre, A.~Yadav, {Performance analysis of NSL-KDD dataset using ANN}, in:
  2015 International Conference on Signal Processing and Communication
  Engineering Systems, 2015, pp. 92--96.
\newblock \href {https://doi.org/10.1109/SPACES.2015.7058223}
  {\path{doi:10.1109/SPACES.2015.7058223}}.

\bibitem{ml-insec}
F.~Ceschin, M.~Botacin, A.~Bifet, B.~Pfahringer, L.~S. Oliveira, H.~M. Gomes,
  A.~Gr\'{e}gio, \href{https://doi.org/10.1145/3617897}{{Machine Learning (In)
  Security: A Stream of Problems}}, Digital Threats 5~(1) (mar 2024).
\newblock \href {https://doi.org/10.1145/3617897} {\path{doi:10.1145/3617897}}.
\newline\urlprefix\url{https://doi.org/10.1145/3617897}

\bibitem{pragmatic-ml}
G.~Apruzzese, P.~Laskov, J.~Schneider,
  \href{https://doi.ieeecomputersociety.org/10.1109/EuroSP57164.2023.00042}{{SoK:
  Pragmatic Assessment of Machine Learning for Network Intrusion Detection}},
  in: 2023 IEEE 8th European Symposium on Security and Privacy (EuroS\&P), IEEE
  Computer Society, Los Alamitos, CA, USA, 2023, pp. 592--614.
\newblock \href {https://doi.org/10.1109/EuroSP57164.2023.00042}
  {\path{doi:10.1109/EuroSP57164.2023.00042}}.
\newline\urlprefix\url{https://doi.ieeecomputersociety.org/10.1109/EuroSP57164.2023.00042}

\bibitem{kalakoti}
R.~Kalakoti, H.~Bahsi, S.~Nõmm, {Improving IoT Security With Explainable AI:
  Quantitative Evaluation of Explainability for IoT Botnet Detection}, IEEE
  Internet of Things Journal 11~(10) (2024) 18237--18254.
\newblock \href {https://doi.org/10.1109/JIOT.2024.3360626}
  {\path{doi:10.1109/JIOT.2024.3360626}}.

\bibitem{patil}
S.~Patil, V.~Varadarajan, S.~M. Mazhar, A.~Sahibzada, N.~Ahmed, O.~Sinha,
  S.~Kumar, K.~Shaw, K.~Kotecha,
  \href{https://www.mdpi.com/2079-9292/11/19/3079}{{Explainable Artificial
  Intelligence for Intrusion Detection System}}, Electronics 11~(19) (2022).
\newblock \href {https://doi.org/10.3390/electronics11193079}
  {\path{doi:10.3390/electronics11193079}}.
\newline\urlprefix\url{https://www.mdpi.com/2079-9292/11/19/3079}

\bibitem{arreche}
O.~Arreche, T.~R. Guntur, J.~W. Roberts, M.~Abdallah, {E-XAI: Evaluating
  Black-Box Explainable AI Frameworks for Network Intrusion Detection}, IEEE
  Access 12 (2024) 23954--23988.
\newblock \href {https://doi.org/10.1109/ACCESS.2024.3365140}
  {\path{doi:10.1109/ACCESS.2024.3365140}}.

\bibitem{barnard}
P.~Barnard, N.~Marchetti, L.~A. DaSilva, {Robust Network Intrusion Detection
  Through Explainable Artificial Intelligence (XAI)}, IEEE Networking Letters
  4~(3) (2022) 167--171.
\newblock \href {https://doi.org/10.1109/LNET.2022.3186589}
  {\path{doi:10.1109/LNET.2022.3186589}}.

\bibitem{cicids-2017}
I.~Sharafaldin, A.~H. Lashkari, A.~A. Ghorbani, {Toward Generating a New
  Intrusion Detection Dataset and Intrusion Traffic Characterization}, in:
  Proceedings of the 4th International Conference on Information Systems
  Security and Privacy - Volume 1: ICISSP,, INSTICC, SciTePress, 2018, pp.
  108--116.
\newblock \href {https://doi.org/10.5220/0006639801080116}
  {\path{doi:10.5220/0006639801080116}}.

\bibitem{shtayat}
M.~M. Shtayat, M.~K. Hasan, R.~Sulaiman, S.~Islam, A.~U.~R. Khan, {An
  Explainable Ensemble Deep Learning Approach for Intrusion Detection in
  Industrial Internet of Things}, IEEE Access 11 (2023) 115047--115061.
\newblock \href {https://doi.org/10.1109/ACCESS.2023.3323573}
  {\path{doi:10.1109/ACCESS.2023.3323573}}.

\bibitem{doh-explain}
T.~Zebin, S.~Rezvy, Y.~Luo, {An Explainable AI-Based Intrusion Detection System
  for DNS Over HTTPS (DoH) Attacks}, IEEE Transactions on Information Forensics
  and Security 17 (2022) 2339--2349.
\newblock \href {https://doi.org/10.1109/TIFS.2022.3183390}
  {\path{doi:10.1109/TIFS.2022.3183390}}.

\bibitem{ensemble-explain}
C.~Minh, K.~Vermeulen, C.~Lefebvre, P.~Owezarski, W.~Ritchie, {An
  Explainable-by-Design Ensemble Learning System to Detect Unknown Network
  Attacks}, in: 2023 19th International Conference on Network and Service
  Management (CNSM), 2023, pp. 1--9.
\newblock \href {https://doi.org/10.23919/CNSM59352.2023.10327818}
  {\path{doi:10.23919/CNSM59352.2023.10327818}}.

\bibitem{bastian-1}
E.~Alhajjar, P.~Maxwell, N.~Bastian,
  \href{https://www.sciencedirect.com/science/article/pii/S0957417421011507}{{Adversarial
  machine learning in Network Intrusion Detection Systems}}, Expert Systems
  with Applications 186 (2021) 115782.
\newblock \href {https://doi.org/https://doi.org/10.1016/j.eswa.2021.115782}
  {\path{doi:https://doi.org/10.1016/j.eswa.2021.115782}}.
\newline\urlprefix\url{https://www.sciencedirect.com/science/article/pii/S0957417421011507}

\bibitem{bastian-2}
M.~Chalé, B.~Cox, J.~Wier, N.~D. Bastian, {Constrained optimization based
  adversarial example generation for transfer attacks in network intrusion
  detection systems}, Optimization Letters 18 (2021) 2169--2188.
\newblock \href {https://doi.org/https://doi.org/10.1007/s11590-023-02007-7}
  {\path{doi:https://doi.org/10.1007/s11590-023-02007-7}}.

\end{thebibliography}

%

\end{document}